\documentclass[a4paper,11pt]{article}
\pdfoutput=1 

\usepackage{jheppub} 
\usepackage{amsmath,amssymb,amsthm,amscd,graphicx}
\input epsf.sty

\theoremstyle{definition}

\newcommand{\CC}{{\cal C}}

\newcommand{\CH}{{\cal H}}
\newcommand{\CI}{{\cal I}}

\newcommand{\CO}{{\cal O}}

\def\IZ{{\mathbb Z}}
\def\IR{{\mathbb R}}
\def\IC{{\mathbb C}}

\def\IP{{\mathbb P}}

\newcommand{\re}{{\rm e}}
\newcommand{\ri}{\mathsf{i}}
\newcommand{\rd}{{\rm d}}

\newcommand{\mx}{\mathsf{x}}
\newcommand{\my}{\mathsf{y}}
\newcommand{\mm}{\mathsf{p}}
\newcommand{\im}{\mathsf{i}}
\newcommand{\mH}{\mathsf{H}}

\newcommand{\mq}{\mathsf{q}}

\newcommand{\be}{\begin{equation}}
\newcommand{\ee}{\end{equation}}
\newcommand{\ba}{\begin{aligned}}
\newcommand{\ea}{\end{aligned}}
\newcommand{\ben}{\begin{eqnarray}\displaystyle}
\newcommand{\een}{\end{eqnarray}}

\newcommand{\sectiono}[1]{\section{#1}\setcounter{equation}{0}}

\newcommand{\beq}{\begin{equation}}
\newcommand{\eeq}{\end{equation}}

\newdimen\tableauside\tableauside=1.0ex
\newdimen\tableaurule\tableaurule=0.4pt
\newdimen\tableaustep
\def\phantomhrule#1{\hbox{\vbox to0pt{\hrule height\tableaurule width#1\vss}}}
\def\phantomvrule#1{\vbox{\hbox to0pt{\vrule width\tableaurule height#1\hss}}}
\def\sqr{\vbox{%
  \phantomhrule\tableaustep
  \hbox{\phantomvrule\tableaustep\kern\tableaustep\phantomvrule\tableaustep}%
  \hbox{\vbox{\phantomhrule\tableauside}\kern-\tableaurule}}}
\def\squares#1{\hbox{\count0=#1\noindent\loop\sqr
  \advance\count0 by-1 \ifnum\count0>0\repeat}}
\def\tableau#1{\vcenter{\offinterlineskip
  \tableaustep=\tableauside\advance\tableaustep by-\tableaurule
  \kern\normallineskip\hbox
    {\kern\normallineskip\vbox
      {\gettableau#1 0 }%
     \kern\normallineskip\kern\tableaurule}%
  \kern\normallineskip\kern\tableaurule}}
\def\gettableau#1{\ifnum#1=0\let\next=\null\else
\squares{#1}\let\next=\gettableau\fi\next}

\newcommand{\figref}[1]{Fig.~\protect\ref{#1}}

\def\({\left(}
\def\){\right)}

\title{\boldmath Exact quantization conditions for cluster integrable systems}

\author[a,b]{Sebasti\'an Franco,}
\author[c]{Yasuyuki Hatsuda}
\author[c]{and Marcos Mari\~no}
\affiliation[a]{
Physics Department, The City College of the CUNY \\
160 Convent Avenue, New York, NY 10031, USA}

\affiliation[b]{The Graduate School and University Center, The City University of New York  \\
365 Fifth Avenue, New York NY 10016, USA }

\affiliation[c]{D\'epartement de Physique Th\'eorique et section de Math\'ematiques\\
Universit\'e de Gen\`eve, Gen\`eve, CH-1211 Switzerland}

\emailAdd{sfranco@ccny.cuny.edu,Yasuyuki.Hatsuda@unige.ch, Marcos.Marino@unige.ch} 

\preprint{
\begin{flushright}
CCNY-HEP-15-08 
\end{flushright}
}

\abstract{We propose exact quantization conditions for the quantum integrable systems of Goncharov and Kenyon, based on the enumerative geometry 
of the corresponding toric Calabi--Yau manifolds. Our conjecture builds upon 
recent results on the quantization of mirror curves, and generalizes a previous proposal for the quantization of the relativistic Toda lattice. We present explicit tests of our 
conjecture for the integrable systems associated to the resolved $\IC^3/\IZ_5$ and $\IC^3/\IZ_6$ orbifolds.}    

\begin{document}
\maketitle
\flushbottom

\sectiono{Introduction}

One fascinating aspect of string theory and supersymmetric gauge theory is the recurrent appearance of connections to integrable systems. These connections provide in many cases 
exact solutions to string or field theory models. Conversely, developments in string theory and in supersymmetric gauge theories have led to new insights 
and methods in integrability. For example, it has been known for some time \cite{itep,mw} that ${\cal N}=2$ supersymmetric Yang--Mills theories are deeply related 
to classical integrable systems of the Toda type. Progress in instanton counting in these gauge theories \cite{nn} eventually led to quantization conditions for the 
Toda lattice \cite{ns} which are in many ways preferable to the solution obtained with the conventional methods of integrable systems.  

Relativistic systems of the Toda type \cite{ruijs} are connected to gauge theories in five dimensions \cite{nek5d}, and also to topological strings on 
toric Calabi--Yau (CY) manifolds. For example, the relativistic Toda lattice with $N$ particles is closely related to topological string theory on 
a certain $A_{N-1}$ fibration over $\IP^1$. Based on recent progress on the quantization 
of mirror curves for toric CY manifolds \cite{km, hw, ghm, gkmr,wzh,cgm}, exact quantization conditions for the relativistic Toda lattice have been recently 
proposed in \cite{hm}. The starting point for these conditions are the perturbative results of \cite{ns}, but one has to add in addition non-perturbative corrections in order to 
obtain the correct answer. 

The relativistic Toda lattice is realized by a particular family of toric CY manifolds. A natural question is then whether one can associate a quantum integrable system to {\it any} toric CY. 
This was answered in the affirmative in the beautiful work of Goncharov and Kenyon \cite{gk}, who showed that, given any Newton polygon in two dimensions, one can 
construct in a purely combinatorial fashion an integrable system associated to it. Since toric CY manifolds are also labeled by these Newton polygons, the construction 
of \cite{gk} allows us to associate an integrable system to an arbitrary toric CY. In particular, as shown in \cite{efs}, the relativistic Toda lattice is 
recovered when the CY is the appropriate $A_{N-1}$ fibration over $\IP^1$. We will refer to the integrable systems constructed in \cite{gk} as GK, or cluster integrable systems. 

Once a quantum integrable system has been associated to an arbitrary toric CY, a second natural question is whether one can solve for its spectrum with the tools of topological string theory. 
The exact solution proposed in \cite{hm} for the relativistic Toda lattice can be generalized to any toric CY, and it was already proposed in \cite{hm} that this generalization should solve the corresponding GK system. 
In this paper, building on \cite{hm} we present a conjectural, exact quantization condition for the GK integrable system, in terms of the topological string free energy of the 
corresponding toric CY. More precisely, our quantization condition determines \textit{all} eigenvalues
of the mutually commuting Hamiltonians for \textit{any} real $\hbar$ (so that the underlying Hamiltonians are self-adjoint), and it involves 
the so-called Nekrasov--Shatashvili (NS) limit of the refined topological string free energy, which is determined by the refined BPS invariants of the CY. In this way, the solution to the 
GK system gets related to the enumerative geometry of the CY\footnote{In this paper we will restrict ourselves to toric CY manifolds with a mirror curve of genus greater or equal than one, since the definition 
of the integrable system in the case of genus zero is more subtle and requires a limiting procedure.}.

 When the CY corresponds to a Newton polygon with a single inner point, so that the mirror curve 
has genus one, the GK integrable system consists of a single Hamiltonian. It turns out that this Hamiltonian is precisely 
the functional difference operator associated to the mirror curve in \cite{adkmv,acdkv}. In this case, the exact quantization condition for the GK Hamiltonian proposed in this paper 
agrees with the proposal in \cite{ghm, wzh} for genus one mirror curves, which has been extensively tested. Our conjecture recovers as well the exact solution for the 
relativistic Toda lattice proposed in \cite{hm}. Therefore, in order to test our proposal for the general GK system, we focus in this paper 
on the GK integrable systems associated to the resolved $\IC^3/\IZ_5$ and $\IC^3/\IZ_6$ orbifolds. These 
examples have two commuting Hamiltonians, and mirror curves of genus two. They are more complicated than the genus one models studied in 
\cite{ghm,gkmr,wzh}, but they are still amenable to a detailed study. As in the $N=3$ relativistic Toda lattice considered in \cite{hm}, we
 show that the numerical diagonalization of the Hamiltonians 
agrees with high precision with the predictions of our exact quantization conditions. 

In \cite{cgm}, a different quantization scheme for higher genus mirror curves was proposed. This scheme is based on quantizing the mirror curve directly, and leads 
to a single quantization condition, which can be obtained as the zero locus of a (generalized) spectral determinant in moduli space. The advantage of the quantization method 
of \cite{cgm} is twofold: one works with one-dimensional operators, and one can in fact reconstruct the topological string free energy with the spectral traces of these operators. 
The approach of \cite{cgm} can be also regarded as providing a solution to the quantum Baxter equation of the GK system (after separation of variables), and in this paper we show 
explicitly, in the example of the resolved $\IC^3/\IZ_5$ orbifold, how the solution of the GK integrable system is encoded in the generalized spectral determinant of \cite{cgm}. 

This paper is organized as follows. In section \ref{section_intro_GK} we present a lightning review of GK integrable systems. In section \ref{section_cluster_ZN} we construct the GK integrable systems associated to an infinite class of $\IC^3/\IZ_N$ orbifolds. In section \ref{section_exact_quantization} we present our general conjecture giving the exact quantization conditions for the quantum GK integrable system. In section \ref{section_Z5_Z6} we illustrate the conjecture with a detailed analysis of the resolved $\IC^3/\IZ_5$ and $\IC^3/\IZ_6$ orbifolds. In section \ref{section_conclusions} we conclude and mention some problems for the future. 

\sectiono{Cluster integrable systems}

\label{section_intro_GK}

In \cite{gk}, Goncharov and Kenyon introduced the infinite class of {\it cluster integrable systems}. In this section we present a brief review of their construction. 
The GK correspondence associates an integrable system to every convex Newton polygon $N$ in $\mathbb{Z}^2$. Such polygons can be regarded as toric diagrams of toric CY 3-folds. We denote the CY associated to $N$ by $X_N$.

Cluster integrable systems can be defined in terms of the dimer model(s), namely bipartite graphs embedded in a 2-torus, corresponding to $X_N$. Generically, more than one dimer model are associated to a given $X_N$. These multiple dimer models are related by local square moves, which translate into cluster transformations connecting different patches of the phase space of the underlying integrable system. 

\begin{figure}[!ht]
\begin{center}
\includegraphics[width=11.5cm]{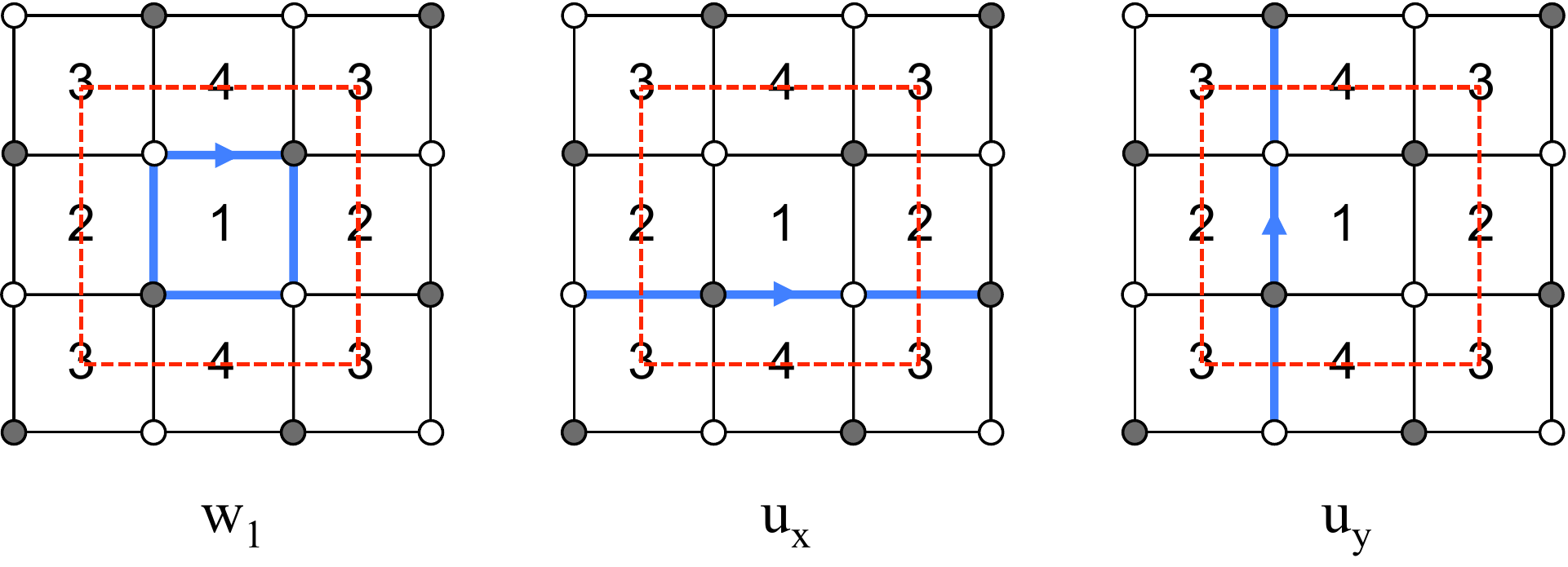}
\caption{An example of a dimer showing the $w_1$, $u_x$ and $u_y$ basic paths. The red square indicates the unit cell.}
\label{sample_wxy_paths}
\end{center}
\end{figure}

The construction of a cluster integrable system can be summarized as follows. The dynamical variables of the integrable system are given by oriented loops on the dimer model. A natural basis for such loops is given by the cycles $w_i$ that go clockwise around each face ($i=1,\ldots,N_F$) and a pair of cycles $u_x$ and $u_y$ wrapping the two fundamental directions of the 2-torus.\footnote{One of the $w_i$'s is redundant, since they are subject to the constraint $\prod_{i=1}^{N_F} w_i=1$.}  \figref{sample_wxy_paths} illustrates these paths in a simple example. The number of faces in the dimer is $N_F=2A_N$, with $A_N$ the area of $N$. We thus conclude that the number of independent paths is equal to $2A_N+1$. It is possible and often convenient, as in the examples presented in section \ref{section_cluster_ZN}, to consider a different basis of paths.

The Poisson bracket between any pair of oriented paths $u$ and $v$ is given by
\beq
\{u,v\}  = ( \langle u,v \rangle +\epsilon_{u,v }) \, u \, v \, ,
\label{Poisson_brackets}
\eeq
where $\epsilon_{u,v}$ is the number of edges over which $u$ and $v$ overlap, counted with orientation, and $\langle u,v \rangle$ is the intersection number in homology between the two paths. For the simple basis discussed above, this implies
\beq
\begin{array}{ccl}
\{w_i,w_j\} & = & \epsilon_{w_i,w_j} \, w_i w_j \\[.15cm]
\{u_x,u_y\} & = & (\langle u_x,u_y \rangle +\epsilon_{u_x,u_y}) \, u_x u_y \\[.15cm]
\{u_a,w_i\} & = & \epsilon_{u_a,w_i} \, u_a w_i
\end{array}
\eeq

The Newton polygon $N$ emerges from the dimer combinatorially. Every point in $N$ corresponds to a collection of {\it perfect matchings}. A perfect matching is a collection of edges in the dimer such that every node is the endpoint of exactly one edge in the perfect matching. Remarkably, it is straightforward to determine the perfect matchings of a dimer by computing the determinant of the Kasteleyn matrix, which is an adjacency matrix of the graph (see e.g. \cite{Franco:2005rj}). We will present an explicit example of the Kasteleyn matrix in section \ref{section_cluster_ZN}. 

The next step is to translate perfect matchings into loops, since the integrable system is formulated in terms of them. When doing so, edge weights are canonically oriented from white to black vertices and every perfect matching is associated a weight given by the product of the weights of all edges it contains. Each perfect matching is then mapped into a loop by subtracting a reference perfect matching $p_0$. The orientations of all edges in the subtracted perfect matching are reversed, which corresponds to inverting its edge weights. 

To every point in $N$, we associate the sum the loop contributions coming from all the corresponding perfect matchings. GK showed that the Poisson brackets in (\ref{Poisson_brackets}) give rise to an integrable system whose integrals of motion are given by:
\begin{itemize}
\item{\bf Hamiltonians:} they are in one-to-one correspondence with strictly internal points in $N$. 
\item{\bf Casimirs:} they are defined as the ratio between contributions coming from consecutive points on the boundary of $N$.
\end{itemize}
Denoting $I_N$ the number of internal points in the Newton polygon and $B_N$ the number of points on its boundary, we see that there are $I_N$ Hamiltonians and $B_N-1$ independent Casimirs. It is straightforward to understand integrability at the level of counting integrals of motion. Using the Casimirs to solve for some of the loops in terms of others, the dimensionality of the phase space is equal to $2 A_N-B_N+2$. By Pick's theorem, this is equal to $2I_N$, namely to twice the number of Hamiltonians. We conclude the system is integrable.\footnote{Different choices of $p_0$ result in shifts of the Newton polygon. Such shifts do not affect the Casimirs, which are defined as differences of contributions from pairs of points on the boundary of $N$. However, they do modify the would-be Hamiltonian(s). It is only for special choices of $p_0$ that the construction we described gives rise to an integrable system. GK proved that it is always possible to chose $p_0$ such that this is the case. In the examples of section \ref{section_cluster_ZN}, we will pick $p_0$ appropriately.}


The spectral curve of the integrable system coincides with the mirror curve $\Sigma_N$ for the CY and is given by
\beq
P(x,y)=\sum_{(n_x,n_y)\in N} c_{n_x,n_y} x^{n_x} y^{n_y}=0 \, ,
\eeq 
where $x,y \in \mathbb{C}$. The coefficients $c_{n_x,n_y}$ are the loop contributions constructed above. Two of the coefficients can be absorbed by rescaling $x$ and $y$. This freedom can be exploited to eliminate two of the Casimirs, leaving us with $B_N-3$ independent ones.

The classical integrable system can be quantized by expressing paths in the dimer as exponentials of linear combinations of Heisenberg operators satisfying canonical commutation relations. 
Ordering ambiguities are resolved by using Weyl quantization, which is the most natural prescription when dealing with exponentiated Heisenberg operators, and leads 
automatically to self-adjoint operators. This is also the prescription appearing in the quantization of cluster algebras, as explained in for example section 4.3 of \cite{bz}. 
Explicit examples of this procedure are presented in section \ref{section_Z5_Z6}.

Interestingly, an alternative formulation of cluster integrable systems based on words in the Weyl group of the loop group was developed by Fock and Marshakov \cite{fm}. 

\paragraph{Connection to gauge theory.}
Before closing this section, it is interesting to mention the connection between cluster integrable systems and two different classes of gauge theories. First, the dimer models discussed above are precisely the {\it brane tilings} encoding the $4d$ $\mathcal{N}=1$ quiver gauge theories on D3-branes probing $X_N$ \cite{Franco:2005rj,Franco:2005sm}. On the other hand, in some cases,\footnote{Not every Newton polygon has a $5d$ gauge theory interpretation. For example, for the $\IC^3/\IZ_N$ orbifolds considered in sections \ref{section_cluster_ZN} to \ref{section_Z5_Z6}, only those for even $N$ are associated to $5d$ gauge theories.} 
$\Sigma_N$ can be regarded as the analogue of the Seiberg-Witten curve for a $5d$ $\mathcal{N}=1$ gauge theory compactified on a circle \cite{nek5d}. In the $4d$ limit of such a theory we obtain an $\mathcal{N}=2$ theory. The usual integrable system associated to this gauge theory, whose spectral curve is equal to the Seiberg-Witten curve \cite{itep,mw}, is obtained by taking a non-relativistic limit of the corresponding cluster integrable system. We will not pursue the connections to gauge theories any further in this paper. 

\sectiono{Cluster integrable systems for $\mathbb{C}^3/\mathbb{Z}_N$}

\label{section_cluster_ZN}

\begin{figure}[!ht]
\begin{center}
\includegraphics[width=12cm]{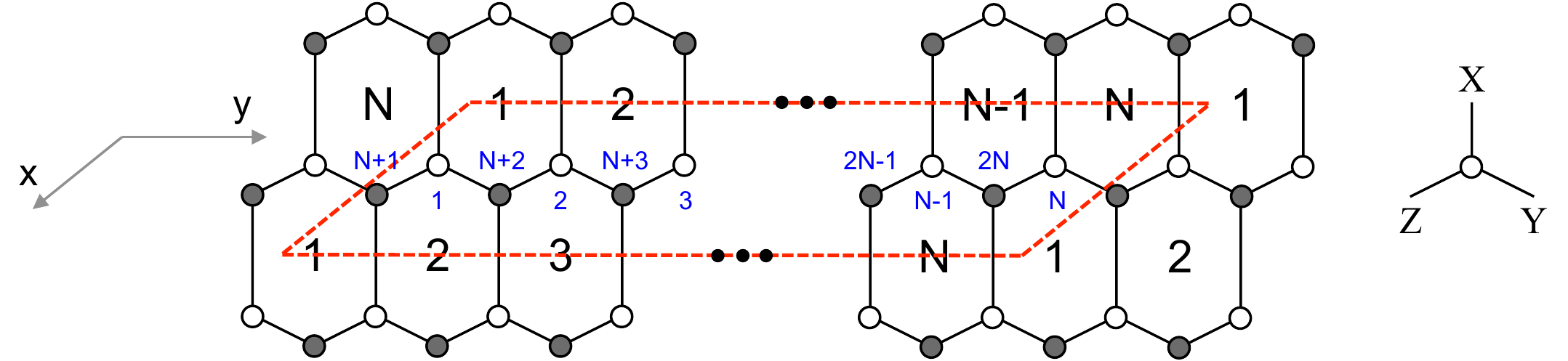}
\caption{Dimer model for $\mathbb{C}^3/\mathbb{Z}_N$. Notice that, for later convenience, we have chosen the $y$ direction to be horizontal. The three edge directions originate from the three complex planes in $\mathbb{C}^3$, as shown on the right. Nodes are numbered in blue for defining the Kasteleyn matrix.}
\label{dimer_ZN}
\end{center}
\end{figure}

In this section we construct the cluster integrable system for an infinite family of orbifolds. Let us consider $\mathbb{C}^3/\mathbb{Z}_N$ orbifolds with geometric action on the three complex planes $(X,Y,Z)\to (\alpha X,\alpha Y, \alpha^{-2}Z)$, with $\alpha^N=1$. The dimer models for these geometries take the general form shown in \figref{dimer_ZN}. The geometric action of the orbifold determines how the $N$ hexagonal faces of the dimer are glued together (see e.g. \cite{Davey:2010px}).

The $N\times N$ Kasteleyn matrix is given by
{\small
\beq
K=\left(\begin{array}{c|cccccc}
& \ \ \ \ N+1 \ \ \ \ & \ \ \ \ N+2 \ \ \ \ & \ \ \ \ N+3 \ \ \ \ & \ \ \ \ \cdots \ \ \ \ & \ \ \ \ 2N-1 \ \ \ \ & \ \ \ \ 2N \ \ \ \ \\ \hline
1 \ \ & Z_{2,N} & Y_{1,2} & 0 & 0 & 0 & X_{N,1} \, x\, y \\
2 \ \ &  X_{12} \, x & Z_{3,1} & Y_{23} & 0 & 0 & 0 \\
3 \ \ & 0 & X_{23} \, x & Z_{4,2} & \ddots  & 0 & 0 \\
\vdots \ \ & 0 & 0 & \ddots & \ddots & Y_{N-2,N-1} & 0 \\
N-1 \ \ & 0 & 0 & 0 & X_{N-2,N-1} \, x & Z_{N,N-2} & Y_{N-1,N} \\
N \ \ & Y_{N,1} \, y^{-1} & 0 & 0 & 0 & X_{N-1,N} \, x & Z_{1,N-1}
\end{array}\right)
\label{K_matrix}
\eeq}
where rows and columns are indexed by white and black nodes in the dimer, respectively. The edges are labeled $X$, $Y$ and $Z$ according to their origin in $\mathbb{C}^3$. Their subindices indicate the pair of faces they separate, with the convention of going clockwise around white nodes and counterclockwise around black nodes.

\begin{figure}[!ht]
\begin{center}
\includegraphics[width=9.5cm]{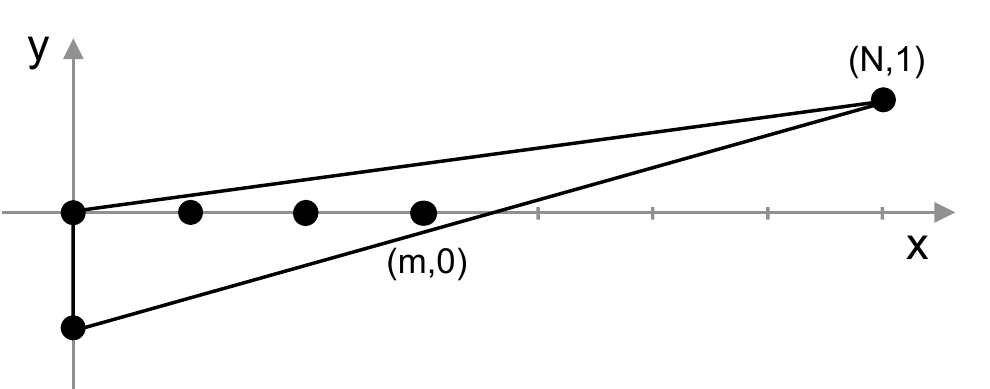}
\caption{Newton polygon for $\mathbb{C}^3/\mathbb{Z}_N$. Here $m=[N/2]$.}
\label{toric_ZN}
\end{center}
\end{figure}

From (\ref{K_matrix}), we obtain the Newton polygon shown in \figref{toric_ZN}. We denote $m=[N/2]$. Notice that for even $N$, the points 
$(0,-1)$, $(m,0)$ and $(N,1)$ become colinear. This is simply a manifestation of the fact that in this case the orbifold becomes $\mathbb{C}^3/(\mathbb{Z}_2\times \mathbb{Z}_m)$, which is also known as the cone over $Y^{m,m}$, a member of the infinite class of $Y^{p,q}$ geometries \cite{Gauntlett:2004zh,Gauntlett:2004yd,Martelli:2004wu,Benvenuti:2004dy}. A general construction of the integrable systems for all $Y^{m,m}$ was given in \cite{efs}. Below we will introduce an even simpler realization.

\subsection{The integrable systems}

All the perfect matchings are easily constructed by taking the determinant of the Kasteleyn matrix (\ref{K_matrix}). In order to convert them into paths, we take the reference perfect matching $p_0=\{Z_{2,N},Z_{3,1},\ldots, Z_{1,N-1}\}$, which corresponds to the point at the origin in \figref{toric_ZN}.

\begin{figure}[!ht]
\begin{center}
\includegraphics[width=5cm]{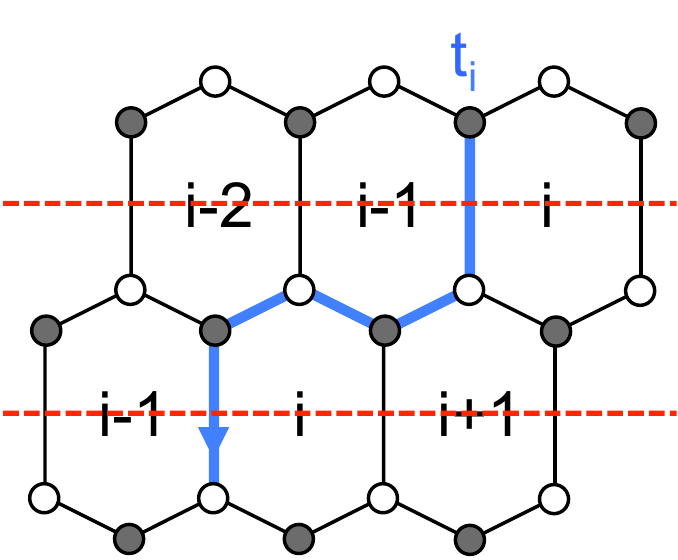}
\caption{Basic paths $t_i$, $i=1,\ldots,N$. The indices are understood mod $N$.}
\label{ti}
\end{center}
\end{figure}

\begin{figure}[!ht]
\begin{center}
\includegraphics[width=8cm]{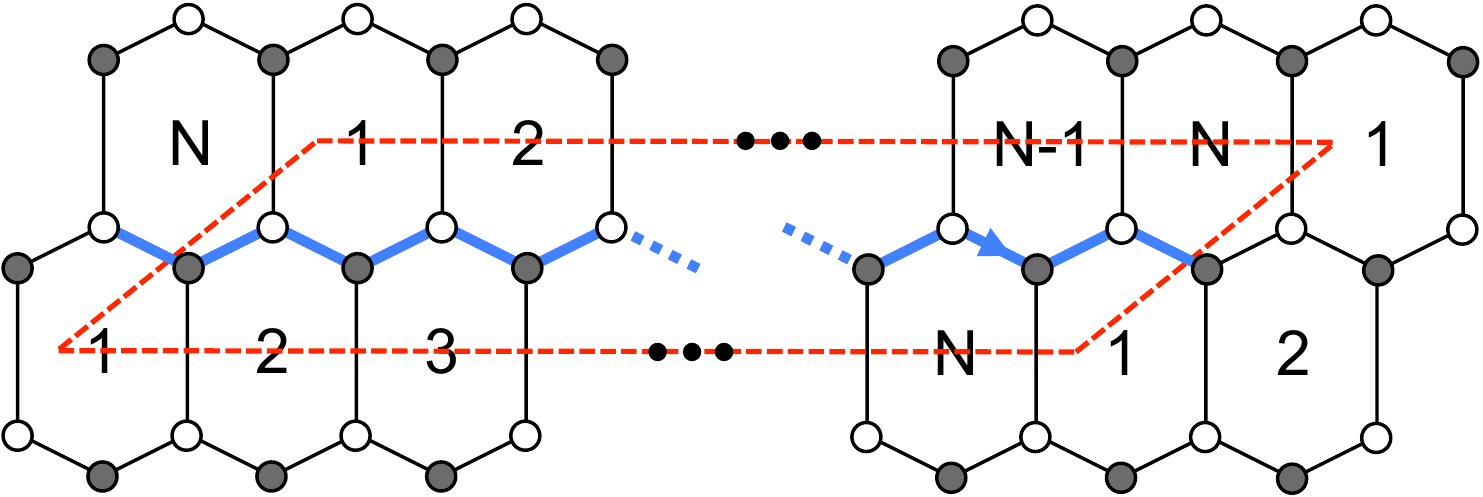}
\caption{The additional path $t_*$ in the basis for $\mathbb{C}^3/\mathbb{Z}_N$.}
\label{t*_ZN}
\end{center}
\end{figure}

Following the general discussion in section \ref{section_intro_GK}, the basis of paths for $\mathbb{C}^3/\mathbb{Z}_N$ contains $N+1$ independent variables. One possibility is to consider the basis of independent faces and fundamental torus cycles: $\{w_i,u_x,u_y \}$, $i=1,\ldots,N-1$. However, as we will see shortly, it is much more convenient to adopt a basis consisting of $N$ paths $t_i$ defined as in \figref{ti}, together with the additional zig-zag path $t_*$ shown in \figref{t*_ZN}. This basis not only trivializes the construction of the integrable system for arbitrary $N$, but also considerably improves the convergence of numerical computations. The only non-vanishing Poisson brackets between these paths are
\beq
\begin{array}{ccr}
\{t_i,t_{i+1}\} & = & t_i t_{i+1}, \\
\end{array}
\label{PB_ti}
\eeq
where indices are understood mod $N$.

The Hamiltonians are simply given by the compact expression
\beq
H_n=\sum_{(i_1,\ldots,i_n)} t_{i_1} \ldots t_{i_n} \, ,
\label{H_n}
\eeq
$n=1,\ldots,m$. The sums run over sets $(i_1,\ldots,i_n)$ such that the corresponding $t_i$ paths do not overlap at any point. The multiplicity, which follows from the Kasteleyn matrix, of perfect matchings contributing to each Hamiltonian is nicely reproduced by the combinatorics of these non-intersecting paths. For even $N$, $H_m$ is actually a Casimir, as it is explained below.

All the orbifolds have two Casimir operators
\beq
\begin{array}{cl}
C_1 & =t_* ,\\[.4cm]
C_2 & = (t_1 \ldots t_N) \, t_*\,.
\end{array}
\label{C1_C2}
\eeq
For odd $N$, there are three points on the boundary of the Newton polygon, so $C_1$ and $C_2$ are the only Casimirs. Both of them can be set to 1 by rescaling $x$ and $y$, as explained in section \ref{section_intro_GK}. For even $N$, there is a fourth point on the boundary of $N$, which results in an additional Casimir
\beq
C_3 = H_m \, t_* \,.
\label{C3}
\eeq
Due to the rescaling of $x$ and $y$, there is only one non-trivial combination of these three Casimirs, which can be identified 
with the single mass parameter of $\Sigma_N$ that is present for even $N$. 

Since $t_*$ is a zig-zag path with vanishing Poisson brackets with all other paths in the basis, it is convenient to define simplified Casimir operators
\beq
\begin{array}{cl}
\tilde{C}_1 & =t_* , \\[.4cm]
\tilde{C}_2 & = t_1 \ldots t_N,  \\[.4cm]
\tilde{C}_3 & = H_m \,.
\end{array}
\label{tilde_Cs}
\eeq
As mentioned earlier, $\tilde{C}_3$ is only a Casimir for even $N$. Otherwise, it is a standard Hamiltonian. Using the Poisson brackets in (\ref{PB_ti}), it is straightforward to show that all the Hamiltonians and Casimirs have the correct commutation relations.

For later analysis, it is convenient to express the variables $t_i$ in terms of $m+1$ pairs of canonical variables.
For $N=2m+1$, a possible choice is
\beq
\begin{array}{rclcccrclcrclcrcl}
t_1&=&\re^{q_m-q_{m+1}}, & \ \ & t_2=\re^{p_m}, & \ \ & t_3 & = & \re^{q_{m-1}-q_m}, & \ \ & t_4 & =& \re^{p_{m-1}},
& \ \ & t_5 & = & \re^{q_{m-2}-q_{m-1}}, \\
t_6 &= & \re^{p_{m-2}}, & \ \ & \dots & \ \ & t_{2m-1} & = & \re^{q_{m+1}-q_1}, & \ \ & t_{2m} & = & \re^{p_1}, & \ \ 
& t_{2m+1}& =& \re^{p_{m+1}+q_{m+1}-q_1},
\end{array}
\label{eq:t-para-odd}
\eeq
where the Poisson brackets of the canonical variables are
\be
\{ q_i , p_j \}=\delta_{ij}, \qquad i,j=1, \cdots, m+1. 
\ee
It is straightforward to verify that \eqref{eq:t-para-odd} indeed gives rise to the Poisson brackets \eqref{PB_ti}.
Moreover the two Casimirs $C_1$ and $C_2$ can be set to one in the center of mass frame.
Then, the explicit form of the first Hamiltonian is 
\be
H_1^{\mathbb{C}^3/\mathbb{Z}_{2m+1}}=\sum_{i=1}^{2m+1} t_i
=\sum_{i=1}^m (\re^{p_i}+\re^{q_i-q_{i+1}})+\re^{p_{m+1}+q_{m+1}-q_1}.
\ee

A similar parameterization is possible for $N=2m$. In this case, as mentioned above, there are three Casimirs. Two them are set to one in the center of mass frame. The third Casimir gives rise to a true independent parameter that we associate to a new variable $R$. From a $5d$ gauge theory viewpoint, $R$ is the radius of the compactification circle. The following parameterization is convenient for studying the $4d$ limit
\beq
\begin{array}{rclcccrclcrclcrcl}
t_1 & = & R^2 \re^{q_1-q_2}, & \ \ & t_2=\re^{p_1},& \ \ & t_3 & = & R^2 \re^{q_m-q_1}, & \ \ & t_4 & = & \re^{p_m}, & \ \ 
& t_5 & = & R^2 \re^{q_{m-1}-q_m}, \\
t_6 & = & \re^{p_{m-1}}, & \ \ & \dots & \ \ & t_{2m-2} & = & \re^{p_3}, & \ \ & t_{2m-1} & = & R^2 \re^{q_2-q_3}, & \ \ & t_m & = & \re^{p_2}.
\end{array}
\label{eq:t-para-even}
\eeq
In this parameterization, the first Hamiltonian in \eqref{H_n} becomes
\be
H_1^{\mathbb{C}^3/\mathbb{Z}_{2m}}=\sum_{i=1}^{2m} t_i=\sum_{i=1}^m (\re^{p_i}+R^2\re^{q_i-q_{i+1}}) \, , 
\label{eq:H1}
\ee
where we impose periodic boundary conditions: $q_{m+1}\equiv q_1$.
This is very reminiscent of the Hamiltonian for the $m$-site relativistic periodic Toda lattice.
Similarly, the last Hamiltonian is given by
\be
H_m^{\mathbb{C}^3/\mathbb{Z}_{2m}}=t_1 t_3 \cdots t_{2m-1}+t_2 t_4 \cdots t_{2m}=R^{2m}+\exp \biggl[\, \sum_{i=1}^m p_i \biggr].
\ee
The $4d$ limit is obtained by taking $R \to 0$ while scaling $p_n = R \, \widetilde{p}_n$. In this limit, \eqref{eq:H1} reduces to the non-relativistic periodic Toda Hamiltonian.
In fact, it was shown in \cite{efs} that all the cluster integrable systems associated with the $Y^{m,n}$ geometries reduce
to the same non-relativistic Toda lattice of $m$ particles.
The Hamiltonian \eqref{eq:H1} explicitly shows this fact for the case of $Y^{m,m}$.

Finally, the quantization of the system is straightforward.
Let $(\mq_i, \mm_j)$ be quantum mechanical canonical operators satisfying
\be
[\mq_i, \mm_j]=\ri \hbar \delta_{ij},
\ee
or equivalently
\be
\re^{\mq_i} \re^{\mm_j}=q^{\delta_{ij}} \re^{\mm_j} \re^{\mq_i},\qquad q=\re^{\ri \hbar}.
\ee
Functions on phase space are promoted to quantum operators by using Weyl quantization. In this way, the exponentials of linear combinations of 
position and momenta become self-adjoint quantum operators
\be
\exp \left[ \sum_{i} (a_i p_i+b_i q_i) \right] \to \exp \left[ \sum_{i} (a_i \mm_i+b_i \mq_i) \right].
\ee
This prescription guarantees the hermiticity of the quantum Hamiltonians.

\subsection{Explicit examples}

For illustration, below we collect a few explicit examples of the general expressions (\ref{H_n}) and (\ref{tilde_Cs}). In section \ref{section_Z5_Z6}, we will focus on $\mathbb{C}^3/\mathbb{Z}_5$ and $\mathbb{C}^3/\mathbb{Z}_6$ to test our conjecture regarding the exact quantization conditions for cluster integrable systems.


\medskip

\noindent $\bullet$ $\mathbb{C}^3/\mathbb{Z}_3$.
In this case, there is a single Hamiltonian, and the spectral problem reduces to a one-particle problem
in the center of mass frame.
In the parameterization \eqref{eq:t-para-odd}, the Hamiltonian is given by
\beq
\begin{array}{cl}
H = & t_1 + t_2 + t_3=\re^{\mq_1-\mq_2}+\re^{\mm_1}+\re^{\mm_2+\mq_2-\mq_1}.
\end{array}
\eeq
One can choose variables $(\mx, \my)$ in the center of mass frame ($p_1+p_2=0$) in such a way that
\be
\mq_1-\mq_2=\mx, \qquad \mm_1=\my, \qquad [\mx, \my]=\ri \hbar.
\label{eq:COM-Z3}
\ee
Thus we obtain
\be
H = \re^\mx+\re^\my+\re^{-\mx-\my}.
\ee
This is just the Hamiltonian obtained from the quantization of the mirror curve of local $\mathbb{P}^2$.
The corresponding spectral problem was first studied in \cite{hw}. An exact quantization condition was proposed in \cite{ghm}, and later reformulated in \cite{wzh}. 

\vspace{0.3cm}

\noindent $\bullet$ $\mathbb{C}^3/\mathbb{Z}_4$.
In this case, one can use the same reparameterization as \eqref{eq:COM-Z3}.
Then the Hamiltonians are given by
\beq
\label{h-c3z4}
\begin{array}{cl}
H_1 = & t_1 + t_2 + t_3 + t_4 =R^2(\re^{\mx}+\re^{-\mx})+\re^{\my}+\re^{-\my}, \\[.2cm]
H_2 = & \tilde{C}_3 = t_1 t_3 + t_2 t_4=1+R^4. 
\end{array}
\eeq
The first Hamiltonian is the same one that one obtains by quantizing the mirror cure of local $\mathbb{F}_0$ \cite{km,hw,ghm}.
Since the resolved $\mathbb{C}^3/\mathbb{Z}_4$ orbifold or $Y^{2,2}$ geometry is local $\mathbb{F}_2$, the spectral problem for the Hamiltonian $H_1$ should be
solved by the topological string on local $\mathbb{F}_2$.
In fact, as shown in \cite{gkmr,kmz}, the spectral problems for local $\mathbb{F}_0$ and local $\mathbb{F}_2$
are exactly related by an appropriate identification of the parameters on both sides.

\vspace{0.3cm}

\noindent $\bullet$ $\mathbb{C}^3/\mathbb{Z}_5$.
This case will be analyzed in subsection~\ref{section_Z5_example}:
\beq
\label{h-c3z5}
\begin{array}{cl}
H_1 = & t_1 + t_2 + t_3 + t_4 + t_5 ,\\[.2cm]
H_2 = & t_1 t_3 + t_1 t_4 + t_2 t_4+t_2 t_5+t_3 t_5 .
\end{array}
\eeq


\noindent $\bullet$ $\mathbb{C}^3/\mathbb{Z}_6$.
This case will be also analyzed in subsection~\ref{section_Z6_example}:
\beq
\label{h-c3z6}
\begin{array}{cl}
H_1 = & t_1 + t_2 + t_3 + t_4 + t_5 + t_6, \\[.2cm]
H_2 = & t_1 t_3 + t_1 t_4 + t_1 t_5+t_2 t_4+t_2 t_5+t_2 t_6+t_3 t_5+t_3 t_6+t_4 t_6, \\[.2cm]
H_3 = &  \tilde{C}_3 = t_1 t_3 t_5 + t_2 t_4 t_6.
\end{array}
\eeq


\noindent $\bullet$ $\mathbb{C}^3/\mathbb{Z}_7$.
\beq
\begin{array}{cl}
H_1 = & t_1 + t_2 + t_3 + t_4 + t_5 + t_6 + t_7, \\[.2cm]
H_2 = & t_1 t_3 + t_1 t_4 + t_1 t_5+t_1 t_6+t_2 t_4+t_2 t_5+t_2 t_6+t_2 t_7+t_3 t_5+t_3 t_6+t_3 t_7+t_4 t_6 +t_4 t_7+t_5 t_7,
\\[.2cm]
H_3 = & t_1 t_3 t_5 +t_1 t_3 t_6 + t_1 t_4 t_6 + t_2 t_4 t_6 + t_2 t_4 t_7 + t_2 t_5 t_7 + t_3 t_5 t_7 .
\end{array}
\eeq


\noindent $\bullet$ $\mathbb{C}^3/\mathbb{Z}_8$.
\beq
\begin{array}{rl}
H_1 = & t_1 + t_2 + t_3 + t_4 + t_5 + t_6 + t_7 + t_8, \\[.2cm]
H_2 = & t_1 t_3 + t_1 t_4 + t_1 t_5+t_1 t_6+t_1 t_7+
t_2 t_4+t_2 t_5+t_2 t_6+t_2 t_7+t_2 t_8\\
+ & t_3 t_5+t_3 t_6+t_3 t_7+t_3 t_8+
t_4 t_6 +t_4 t_7+t_4 t_8+
t_5 t_7 +t_5 t_8 +
t_6 t_8,
\\[.2cm]
H_3 = & t_1 t_3 t_5 +t_1 t_3 t_6 +t_1 t_3 t_7
+ t_1 t_4 t_6+ t_1 t_4 t_7 + t_1 t_5 t_7
+ t_2 t_4 t_6 + t_2 t_4 t_7 \\ + &t_2 t_4 t_8
+ t_2 t_5 t_7 + t_2 t_5 t_8
+ t_2 t_6 t_8 
+ t_3 t_5 t_7 + t_3 t_5 t_8
+ t_3 t_6 t_8
+ t_4 t_6 t_8 ,
\\[.2cm]
H_4 = & \tilde{C}_3 = t_1 t_3 t_5 t_7+t_2 t_4 t_6 t_8.
\end{array}
\eeq

\sectiono{Exact quantization conditions}

\label{section_exact_quantization}

As we have explained above, one can associate to any two-dimensional Newton polygon a quantum cluster integrable system with $I_N$ mutually commuting 
Hamiltonians $\mH_i$, $i=1, \dots, I_N$. A natural problem is to diagonalize these Hamiltonians simultaneously, and find their eigenvalues $H_1, \cdots, H_{I_N}$, as a function 
of $\hbar$ (which we take to be real, so that the operators are self-adjoint), and of the non-trivial Casimirs. 
In this section, we state a conjectural, exact quantization condition for a general GK integrable system associated to a two-dimensional Newton polygon $N$. 
Our exact quantization condition will be written in terms of the NS 
free energy of the toric CY associated to $N$, $X_N$. Let us thus review how this free energy is constructed. 

The number of K\"ahler parameters of the toric CY $X_N$, which we will denote by $n$, is equal to the second Betti 
number of $X_N$, $b_2(X_N)$. 
As explained in for example \cite{hkp, hkrs, klemm-g2}, the K\"ahler parameters are of two types: there are $g_N$ true moduli 
and $r_N$ mass parameters, so that 
\be
n= b_2(X_N)=g_N + r_N.
\ee
The number of true moduli $g_N$ equals the genus of the mirror curve to $X_N$, and also the number of inner points $I_N$ of the Newton polygon. Therefore, the true moduli 
correspond to the Hamiltonians of the GK integrable system. 
The number of mass parameters $r_N$ equals $B_N-3$, where $B_N$ is the number of points in the boundary of polygon. Therefore, the mass parameters correspond to the 
non-trivial Casimirs. We will denote the K\"ahler parameters by $T_i$, $i=1, \cdots, n$, and we will also use the notation 
\be
Q_i =\re^{-T_i}, \qquad i=1, \dots, n. 
\ee
The so-called {\it refined BPS invariants} $N_{j_L, j_R}^{\bf d}$ of $X_N$ depend on 
the total degree ${\bf d}=(d_1, \cdots, d_n)$, which is a vector of non-negative integers specifying a class in $H_2(X_N)$, 
and on two non-negative half-integers $j_L$, $j_R$. From a physical point of view, the refined BPS invariants are indices, 
counting with signs the number of BPS 
states arising in compactifications of M-theory on $X_N$, and due to M2-branes wrapping curves with class ${\bf d}$ \cite{gv,hiv,ikv}. 
They can be also defined mathematically, through a refinement of the Pandharipande--Thomas invariants of $X_N$ \cite{ckk,no}. 
In practice, these invariants are computed with the refined topological vertex \cite{ak,ikv}, or by using a refined version of the 
holomorphic anomaly equation \cite{kw,hk,hkk}.   

The NS free energy of $X_N$ consists of 
two pieces, which we will call the perturbative and the BPS piece. The perturbative piece is given by 
\be
 F^{\rm NS,\, pert }({\bf Q}, \hbar)= {1\over 6 \hbar} \sum_{i,j,k=1}^n a_{ijk} T_i T_j T_k +\left( { 4 \pi^2 \over \hbar} + \hbar  \right)  \sum_{i=1}^n b_i^{\rm NS} T_i . 
 \ee
 In this equation, the coefficients $a_{ijk}$ are given by the triple intersection numbers of $X_N$ (suitably extended to the non-compact setting), 
 while the coefficients $b_i^{\rm NS}$ can be obtained by using for example the holomorphic anomaly equations of \cite{hk}. 
 The BPS part of the NS free energy involves the refined BPS invariants and it is given by 
 \be
\label{lim-ns}
F^{\rm NS, \, BPS}({\bf Q}, \hbar) = \sum_{j_L, j_R} \sum_{w, {\bf d} } 
N^{{\bf d}}_{j_L, j_R}  \frac{\sin\frac{\hbar w}{2}(2j_L+1)\sin\frac{\hbar w}{2}(2j_R+1)}{2 w^2 \sin^3\frac{\hbar w}{2}} {\bf Q}^{w {\bf d}}, 
\ee
where we have denoted
\be
{\bf Q}^{{\bf d}}= \prod_{i=1}^n Q_i^{d_i}. 
\ee
The NS free energy is a particular limit of the refined topological string free energy of the CY. This limit was first considered in the context 
of instanton counting in supersymmetric gauge theories in \cite{ns}, where it was also conjectured that it captures the quantization conditions 
for the corresponding integrable system. As in \cite{ns}, the parameter $\hbar$ appearing in (\ref{lim-ns}) will be identified with the 
Planck constant of the GK integrable system associated to $X_N$. The expansion of the 
NS free energy around $\hbar=0$ defines functions $F^{\rm NS}_\ell ( {\bf Q})$, 
\be
\label{hbar-ex}
F^{\rm NS}({\bf Q}, \hbar)=\sum_{\ell \ge 0} \hbar^{2\ell-1} F^{\rm NS}_\ell ({\bf Q}). 
\ee
The first term in this expansion is equal to the prepotential of the CY manifold $X_N$, up to a linear term in the K\"ahler parameters: 
\be
F^{\rm NS}_0({\bf Q})= F_0({\bf Q}) + 4 \pi^2 \sum_{i=1}^n b_i^{\rm NS} T_i. 
\ee
We recall that the prepotential has the structure, 
\be
F_0({\bf Q}) ={1\over 6} \sum_{i,j,k=1}^n a_{ijk} T_i T_j T_k +\sum_{{\bf d}}\sum_{w\ge1} {n_0^{\bf d} \over w^3} {\bf Q}^{w {\bf d}}, 
\ee
where 
\be
n_0^{\bf d}= \sum_{j_L, j_R} (2j_L+1)(2j_R+1) N^{{\bf d}}_{j_L, j_R}
\ee
are the genus zero Gopakumar--Vafa invariants \cite{gv}. The prepotential can be 
obtained by using the standard tools of (local) mirror symmetry, see \cite{hkt,ckyz}.

The next ingredient we need in order to write our conjecture is 
the B-field $\boldsymbol{B}$ first considered in \cite{hmmo}, and further used in \cite{ghm,cgm}. This B-field satisfies the following requirement: 
for all ${\bf d}$, $j_L$ and $j_R$ such that the BPS invariant $N^{{\bf d}}_{j_L, j_R} $ is non-vanishing, we have
\be
\label{B-prop}
(-1)^{2j_L + 2 j_R+1}= (-1)^{\boldsymbol{B} \cdot {\bf d}}. 
\ee
For local del Pezzo CY threefolds, the existence of such a B-field was 
established in \cite{hmmo}. A general argument for its existence in more general cases has not been given yet, 
but one can identify it in many higher genus examples \cite{cgm,hm}. 

In our conjecture, as we will see in a moment, the quantization conditions determine values for the K\"ahler parameters of the CY. In order 
to obtain the eigenvalues of the Hamiltonians, we have to relate these to the K\"ahler parameters. 
Let us first note that the Hamiltonians and Casimirs appear as coefficients in the spectral curve. Since the spectral curve is the mirror curve to $X_N$, 
they are naturally identified with moduli in the B-model topological string, i.e. with complex deformation parameters of the mirror manifold. 
In standard mirror symmetry, the moduli space of complex 
deformations of the mirror curve is often parametrized by 
the Batyrev coordinates $z_i$, $i=1, \dots, n$. The relationship between the Batyrev coordinates, 
and the Hamiltonians $H_i$ and Casimirs $C_k$ of the integrable system, can be obtained from the 
vectors of charges specifying the toric geometry of $X_N$ (see for example \cite{hkt,hkp}) and from the explicit form of the spectral curve. It has the structure, 
\be
\label{zHC}
z_i= \prod_{j=1}^{g_N} H_j^{-\CC_{ij}} \prod_{k=1}^{r_N} C_k^{-a_{ik}}, \qquad i=1, \cdots, n. 
\ee
The $n \times g_N$ matrix $\CC_{jk}$ appearing in this expression encodes the relation between 
Batyrev coordinates and true moduli of the mirror curve. The restriction $\CC_{jk}$, $j,k=1, \cdots, g_N$, 
is an invertible matrix which encodes the intersections between the $A$ and the $B$ cycles of the mirror curve \cite{klemm-g2}.
 In the case of the resolved $\IC^3/\IZ_5$ and $\IC^3/\IZ_6$ 
orbifolds, the relationship (\ref{zHC}) is written down in (\ref{z-H1}) and 
in (\ref{eq:moduli-Z6}), respectively. 

In standard mirror symmetry, the Batyrev coordinates are related to the K\"ahler parameters through the so-called mirror map, which has the structure
\be 
\label{cmp}
-T_i= \log(z_i) + \widetilde \Pi_i \left( \boldsymbol{z} \right), \qquad i=1, \cdots, n. 
\ee
In this equation, $ \widetilde \Pi_i \left( \boldsymbol{z} \right)$ is a 
power series in the $z_i$s. As shown in \cite{acdkv}, the ``classical" mirror 
map (\ref{cmp}) can be promoted to a ``quantum" mirror map depending on $\hbar$, 
\be
\label{qmp}
-T_i(\hbar)= \log(z_i) + \widetilde \Pi_i \left( \boldsymbol{z}; \hbar \right), \qquad i=1, \cdots, n, 
\ee
which can be obtained by considering ``quantum" periods of the mirror curve. When $T_i$ (or a linear combination thereof) corresponds to a mass parameter, the mirror map is algebraic, 
and the quantum mirror map equals the classical mirror map \cite{hkrs}. In our exact quantization condition, the Hamiltonians and the Casimirs of the the GK system will enter through 
the quantum mirror map (\ref{qmp}). 

We are now ready to present our conjecture. First of all, we introduce the ``hatted" quantum mirror map, which incorporates the B-field, 
\be
-\widehat T_i(\hbar)= \log(z_i) + \widetilde \Pi_i \left(   \boldsymbol{z}_{\boldsymbol{B}};\hbar\right), \qquad i=1, \dots, n, 
\ee
where
\be
 \boldsymbol{z}_{\boldsymbol{B}}=\left( (-1)^{B_1} z_1, \cdots, (-1)^{B_n} z_n\right), 
 \ee
 i.e. it corresponds to a shift 
 \be
 \log(z_i) \rightarrow \log(z_i) + \pi \ri B_i, \qquad i=1, \cdots, n
 \ee
 in the power series. This shift can be also understood as a consequence of the relation between the 
 Batyrev coordinates $z_i$, and the Hamiltonians and Casimirs. 
 An example of this was analyzed in \cite{hm} in the case of the relativistic Toda lattice, and we 
 will see other examples in the next section. We also introduce the ``hatted" NS free energy, 
 \be
\label{hatfns}
 \widehat F^{\rm NS} ({\bf Q}, \hbar)= F^{\rm NS, \, pert} ({\bf Q}, \hbar)+ F^{\rm NS, \, BPS} (   {\bf{Q}}_{\boldsymbol{B}},\hbar), 
 \ee
 where
 \be
  {\bf{Q}}_{\boldsymbol{B}}=\left( (-1)^{B_1} Q_1, \cdots, (-1)^{B_n} Q_n\right), 
  \ee
  which corresponds to a shift
  \be
  T_i \rightarrow T_i + \pi \ri B_i, \qquad i=1, \cdots, n
  \ee
  in the BPS part of the NS free energy. 
  
We now claim that the exact quantization condition for the cluster integrable system associated to $X_N$ is given by  
\be
\label{EQC}
\sum_{j=1}^{n} \CC_{jk} \left\{ {\partial  \over \partial T_j} \widehat F^{\rm NS}\left (\widehat{\boldsymbol{T}}(\hbar), \hbar \right) +
\frac{\hbar}{2\pi}
 {\partial \over \partial T_j} F^{\rm NS, BPS}\left({2 \pi \over \hbar} \widehat{\boldsymbol{T}}(\hbar)+ \pi \ri \boldsymbol{B}, {4 \pi^2  \over \hbar} \right)\right\}= 2 \pi  \left( n_k+{1\over 2} \right),
 \ee
for $ k=1, \cdots, g_N$. In this equation, the $n_k$ are non-negative integers, labelling the eigenstates of the $g_N$ commuting Hamiltonians, and
 $\CC_{jk}$ is the $n \times g_N$ matrix appearing in (\ref{zHC}). Given values 
for $\hbar$ and for the Casimirs, and given a set of $g_N$ quantum numbers $n_1, \dots, n_{g_N}$, 
the equations (\ref{EQC}) determine values for the K\"ahler parameters. These 
in turn determine the eigenvalues of the Hamiltonians $H_i$, $i=1, \cdots, g_N$, through the quantum mirror map. We note that the quantization condition (\ref{EQC}) was already stated 
in this form in \cite{hm}, for the particular case in which $X_N$ is the CY manifold corresponding to the relativistic Toda lattice. 
In addition, (\ref{EQC}) generalizes previous work on exact quantization conditions for quantum mirror curves \cite{km,hw,ghm,wzh,cgm,hatsuda,hm}. 

Let us make some comments on the conditions (\ref{EQC}). First of all, the first term in the l.h.s. has a perturbative 
expansion in $\hbar$ around $\hbar=0$, which gives in fact the all-orders WKB quantization condition. This is expected from the results of 
\cite{mirmor,acdkv}, where it is shown that the NS free energy resums the WKB expansion, as obtained from the quantized spectral curve. 
In particular, at leading order in $\hbar$ we find the Bohr--Sommerfeld/EBK quantization condition
\be
\label{bs}
\sum_{j=1}^{n} \CC_{jk} \left( {\partial \widehat F_0  \over \partial T_j} + 4 \pi^2 b_j^{\rm NS} \right)= 2 \pi \hbar \left( n_k+{1\over 2} \right), \qquad k=1, \cdots, g_N. 
\ee
This is indeed the correct quantization at leading order: the Liouville torus of the GK integrable system is given by the product of the 
ovals of the mirror curve $\Sigma_N$ \cite{gk}. The action variables can then be identified with the B-periods of the mirror curve which 
are dual to the true moduli. These are precisely the B-periods appearing in the l.h.s. of (\ref{bs}). Note that the 
B-field is however crucial to incorporate the right action variables. 

As first noted in \cite{km}, when $\hbar$ is real the total NS free energy 
is afflicted by an infinitely dense set of poles, as it is obvious from the expression (\ref{lim-ns}). 
These poles are of course also present in the perturbative part of (\ref{EQC}). 
However, the second term in the l.h.s. of (\ref{EQC}), which is non-perturbative in $\hbar$, has poles with opposite residues which {\it cancel} the poles in the perturbative part. 
This cancellation of poles has been proved in \cite{hm}, for general toric CYs, and is closely related to the HMO mechanism in ABJM theory \cite{hmo}. 
After this cancellation has taken place, the l.h.s. of (\ref{EQC}) is a well-defined, formal power series in $Q_i$, $Q_i^{2 \pi/\hbar}$, $i=1, \cdots, n$, which 
is expected to be {\it convergent} in a neighborhood of the large radius point $Q_i=0$. In addition, the values of the K\"ahler parameters which solve (\ref{EQC}) are 
expected to be inside this domain of convergence.

Non-perturbative corrections similar to those appearing in (\ref{EQC}) have 
been extensively discussed in \cite{km,hw,ghm,cgm,hatsuda,hm}. In particular, (\ref{EQC}) generalizes the exact quantization condition 
for genus one mirror curves first proposed in \cite{ghm}, and written in this form in \cite{wzh}. A characteristic feature of (\ref{EQC}) is that 
the perturbative and non-perturbative terms are related by a simultaneous exchange or ``S-duality,"
\be
\label{sdual}
\hbar \leftrightarrow {4 \pi^2 \over \hbar}, \qquad \boldsymbol{T} \leftrightarrow {2 \pi \over \hbar} \boldsymbol{T}.  
\ee
This duality has been emphasized in \cite{hatsuda}, following \cite{wzh}\footnote{The expression of \cite{hmmo} for the total grand potential of the CY, which provides exact 
formulae for the Fredholm determinants associated to quantized mirror curves \cite{ghm}, also display the above symmetry. However, one has to exchange 
in addition the conventional and the NS topological string free energies.}. The 
symmetry (\ref{sdual}) is very natural, since the GK system has a ``modular double" with an $S$-dual 
Planck constant of the same form. What is surprising is that the all-orders WKB quantization condition, 
together with this symmetry, lead to a complete solution for the quantum integrable 
system. 

In view of the symmetry (\ref{sdual}), one would expect that the ``self-dual" case
\be
\hbar=2 \pi
\ee
has some special properties. Indeed, this is the ``maximally supersymmetric case" studied in detail in \cite{ghm} in a closely related context. For 
this special value of $\hbar$, the hatted quantum mirror map becomes the standard mirror map, and the quantization condition reduces to 
\be
\label{SD}
\sum_{j=1}^{n} \CC_{jk} \left\{ -{\partial F_0 \over \partial T_j} +
 \sum_{l=1}^{n} T_l {\partial^2 F_0 \over \partial T_j \partial T_l} +8 \pi^2  b_j^{\rm NS} \right\}= 4 \pi^2 \left( n_k +{1\over 2} \right), \qquad 
k=1, \cdots, g_N.
\ee
Remarkably, this quantization condition only involves the leading part of the NS free energy, i.e. the prepotential of the CY 
$X_N$. As is well-known, this prepotential is encoded in a system of Picard--Fuchs equations 
which can be solved exactly and explicitly. In this way, in the self-dual case, one obtains a simple and efficient formula that determines the eigenvalues of the integrable system. We also note that, in this 
simpler case, the convergence of the power series in the l.h.s. of (\ref{SD}) is a consequence of standard results in mirror symmetry. 
\sectiono{Testing the conjecture}

\label{section_Z5_Z6}

As we have explained in the previous section, when the Newton polygon $N$ has only one inner point (so that the mirror curve has genus one), the quantum integrable system of GK 
has one Hamiltonian, which agrees with the operator 
$\CO_{X_N}$ described in \cite{ghm}. In this case, our conjecture (\ref{EQC}) agrees with the quantization condition of \cite{ghm, wzh} for the eigenvalues of this Hamiltonian. In addition, 
when $X_N$ is the $A_{N-1}$ fibration over $\IP^1$ which leads to the relativistic Toda lattice, (\ref{EQC}) is precisely the conjecture put forward in \cite{hm} for that integrable system. 
Therefore, in order to test (\ref{EQC}), we will focus on geometries leading to integrable systems with two Hamiltonians but different from the $N=3$ relativistic Toda lattice, which was studied in \cite{hm}. The simplest 
examples are the $\IC^3/\IZ_5$ and the $\IC^3/\IZ_6$ resolved orbifolds, which we now analyze in detail.

\subsection{The resolved $\IC^3/\IZ_5$ orbifold}
\label{section_Z5_example}

The Newton polygon for this geometry, as shown in \figref{toric_ZN}, has two inner points and three boundary points. Therefore, 
we have two true moduli and two Hamiltonians, and the mirror 
curve has genus two. In addition, there are no mass parameters, and no non-trivial Casimirs. 
After an appropriate choice of variables, the mirror curve can be written as \cite{cgm}, 
\be
\label{mc-c3z5}
\re^x + \re^y+ \re^{-3x - y} + x_0 \re^{-x} + x_3=0. 
\ee
In this equation, $x_0$ and $x_3$ are the two true moduli. The Batyrev coordinates are 
\be
\label{zxs}
z_1={x_3 \over x_0^3}, \qquad z_2={x_0 \over x_3^2}. 
\ee
This geometry has been studied in certain detail, from the point of view of mirror symmetry, in \cite{ossa,ray,klemm-g2,cgm}. We will follow the conventions of \cite{cgm}. The B-field 
has been determined in \cite{cgm} to be given by 
\be
\label{bc3z5}
{\bf{B}}=(1,0), 
\ee
and we also have that 
\be
\hbar F^{\rm NS,\, pert }({\bf T}, \hbar)= {1\over 15} T_1^3 +{1\over 10} T_1^2 T_2 +{3\over 10} T_1 T_2^2 +{3\over 10}T_2^3- (4 \pi^2 + \hbar^2) \left( {T_1 \over 12} + {T_2 \over 8} \right). 
\ee
The Newton polygon associated to (\ref{mc-c3z5}) is easily mapped to the one in \figref{toric_ZN} by an $SL(2,\IZ)$ transformation. Comparing the equation (\ref{mc-c3z5}) for the mirror curve, to the spectral curve of the integrable system, we find the following identification between the moduli and the 
classical Hamiltonians, 
\be\label{xHs}
x_0=H_2, \qquad x_3=-H_1. 
\ee
The Batyrev coordinates are related to the Hamiltonians by the equations,  
\be
\label{z-H1}
z_1= {H_1 \over H_2^3}, \qquad z_2= {H_2 \over H_1^2}. 
\ee
Note that the minus sign induced by the B-field (\ref{bc3z5}) changes the sign of $z_1$ in the mirror map in accordance with the dictionaries (\ref{xHs}) and (\ref{zxs}). 
Finally, the matrix $\CC_{kj}$ appearing in (\ref{zHC}) is 
\be
\label{C-mat}
\CC=\begin{pmatrix} 3 & -1\\ -1 &2 \end{pmatrix}.
\ee
\begin{figure}[tb]
\begin{center}
\begin{tabular}{cc}
\resizebox{65mm}{!}{\includegraphics{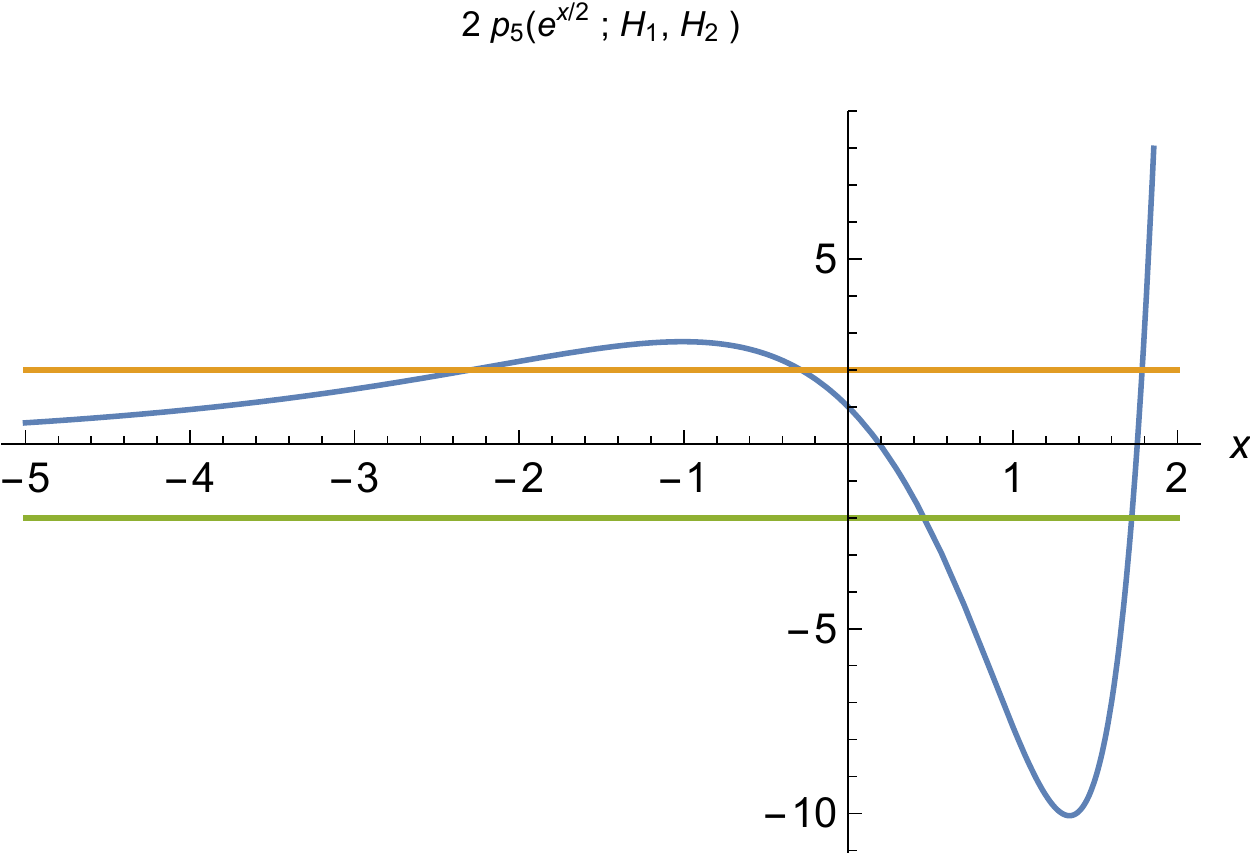}}
\hspace{10mm}
&
\resizebox{55mm}{!}{\includegraphics{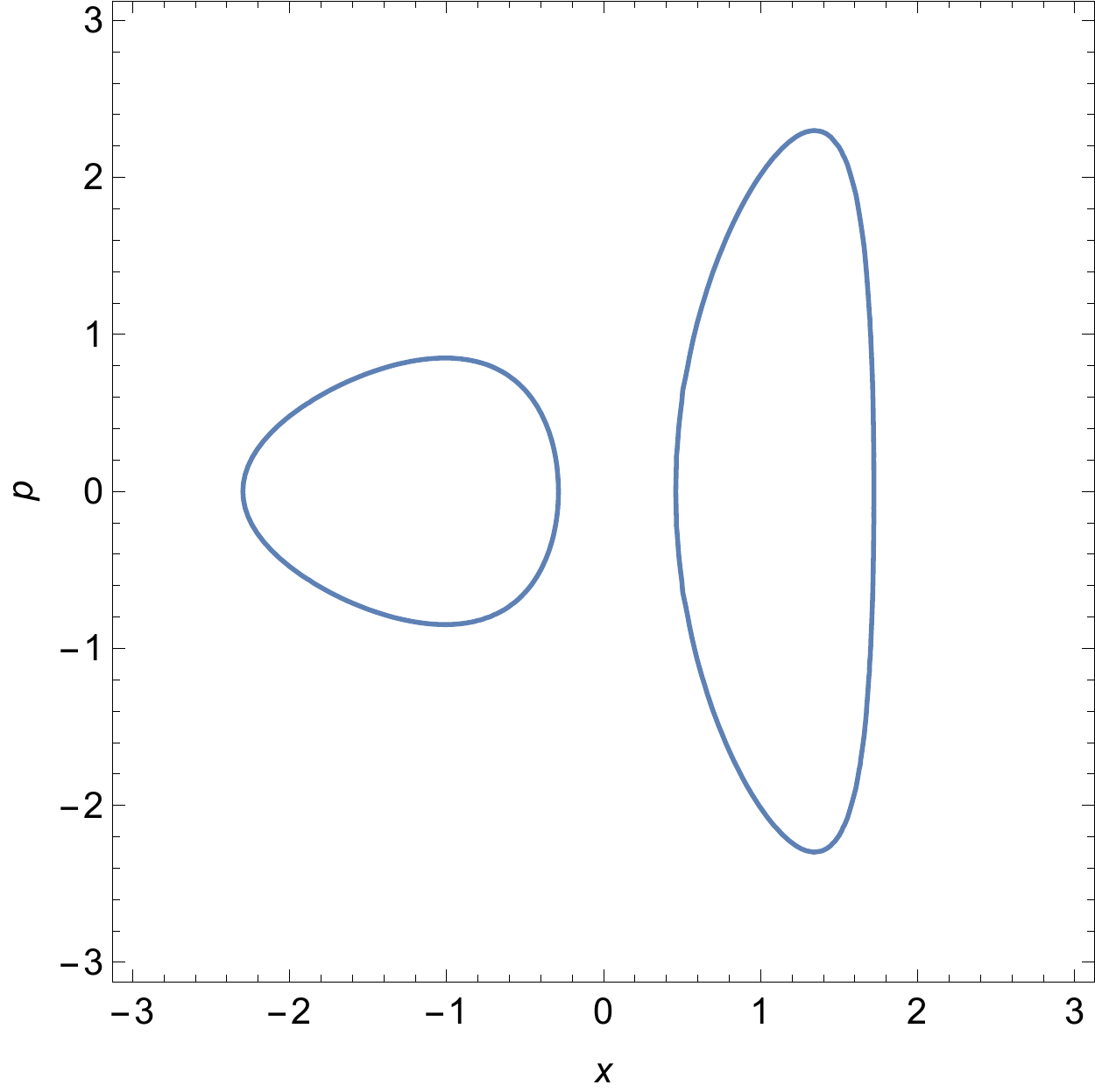}}
\end{tabular}
\end{center}
  \caption{(Left) The function $2p_5(\re^{x/2}; H_1, H_2)$ for $H_1=H_2=7$, as a function of $x$, and the two intervals of instability $\CI_{1,2}$, defined by (\ref{int-ins}).    
  (Right) The corresponding ovals in the $(x,p)$ plane, which are described by (\ref{ovals}).}
\label{tori}
\end{figure}

A first step to understand the integrable system is to consider the Liouville tori and the corresponding periods, which lead to 
the Bohr--Sommerfeld quantization conditions. To do this, 
we perform a change of variables to write the mirror curve in the same form than the spectral curve of the relativistic Toda lattice, 
\be
\label{spec-curve}
\re^{p}+\re^{-p}+2p_5(\re^{x/2}; H_1, H_2) =0, 
\ee
where
\be
2p_5(z;H_1, H_2)=z^5-H_1 z^3+H_2 z. 
\ee
As in the (relativistic) Toda lattice, the Liouville tori correspond to the intervals of instability of this curve, which are defined by 
\be
\label{int-ins}
\left|  p_5(\re^{x/2};H_1, H_2) \right| \ge 1. 
\ee
In this case, there are two intervals of instability in the real $x$-axis, $\CI_{1,2}$, which we show in \figref{tori} for $H_1=H_2=7$. 
These intervals lead to the two ovals of the curve, which correspond to the
two choices of sign in 
\be
\label{ovals}
\cosh p = \pm p_5(\re^{x/2}; H_1, H_2),
\ee
and are also shown in \figref{tori}. The product of these two ovals is the Liouville torus of the integrable system. The action variables 
are then given by period integrals along these ovals, or equivalently, along the intervals of instability:
\be
I_k^{(0)}(H_1,H_2)=2\int_{\mathcal{I}_k}\rd x\,  \cosh^{-1}\left |p_5(\re^{x/2};H_1,H_2)\right|,\qquad
k=1,2.
\ee 
These are periods of the mirror curve, and indeed one confirms that 
\be
I_k^{(0)}(H_1,H_2)=\sum_{j=1}^2 \CC_{jk} \left( {\partial \widehat F_0 \over \partial T_j}+4 \pi^2 b^{\rm NS}_j\right),  \qquad k=1,2.
\ee

We can now proceed to test our conjecture. In order to do this, we have to compute the eigenvalues of the quantum Hamiltonians \eqref{h-c3z5}, 
numerically, and then compare them to the predictions 
obtained from (\ref{EQC}). For the numerical calculation of the eigenvalues, it is convenient to express the Hamiltonians of the GK system in terms of 
exponentiated position and momentum variables, as in \eqref{eq:t-para-odd}. 
In general, if the spectral curve has genus $g_N$, we can introduce $2g_N$ variables $x_i$, $y_i$, $i=1, \cdots, g_N$, satisfying the canonical commutation relations
\be
\{x_i, y_j \}=\delta_{ij}, \qquad i=1, \cdots, g_N. 
\ee
%
%
%
%
%
In the case of the $\IC^3/\IZ_5$, we can write the $t_i$ variables appearing in (\ref{h-c3z5}) as follows, 
\be
t_1=\re^{x_1},\quad t_2=\re^{y_1},\quad t_3=\re^{-x_1+x_2},\quad t_4=\re^{y_2},\quad t_5=\re^{-x_2-y_1-y_2}.
\ee
This is a reparameterization of \eqref{eq:t-para-odd} in the center of mass frame.
The quantum Hamiltonians corresponding to (\ref{h-c3z5}) read, 
\be
\ba
\mH_1&=\re^{\mx_1}+\re^{\my_1}+\re^{-\mx_1+\mx_2}+\re^{\my_2}+\re^{-\mx_2-\my_1-\my_2}, \\
\mH_2&=\re^{\mx_2}+\re^{\mx_1+\my_2}+\re^{\my_1+\my_2}+\re^{-\my_2-\mx_2}+ \re^{-\mx_1-\my_1-\my_2},
\ea
\ee
where $[\mx_i, \my_j]=\ri \hbar \delta_{ij}$. 
They act on the Hilbert space $\CH=L^2(\IR^2)$. 

 Let us now discuss the numerical computation of the spectrum. As in previous numerical studies of quantum integrable systems \cite{matsuyama,isola,hm}, 
 one can choose an appropriate basis for 
the Hilbert space and compute the matrix elements of the Hamiltonians. A useful choice is, 
\be
\label{basis}
\langle x_1, x_2 |m_1, m_2 \rangle=\phi_{m_1} (x_1) \phi_{m_2}(x_2), \qquad m_1, m_2=0, 1, 2, \cdots, 
\ee
where $\phi_m(x)$ are the eigenfunctions of the one-dimensional harmonic oscillator. The eigenvalues of the infinite matrix 
\be
\label{H-matrix}
\langle \ell_1, \ell_2 |\mH_1| m_1, m_2\rangle, 
\ee
give the eigenvalues of $\mH_1$. In practice, one truncates the basis of the Hilbert space to a set of $M$ elements. 
The eigenvalues of the truncated matrix provide an 
approximation to the eigenvalues of $\mH_1$, and they converge to the 
exact eigenvalues as $M$ is increased. The approximate eigenvalues of the second 
Hamiltonian $\mH_2$ can be obtained by calculating its vacuum expectation values in the approximate eigenfunctions 
obtained in the diagonalization of $\mH_1$. However, this procedure is relatively time-consuming. 
As pointed out in \cite{hm} in the case of the relativistic Toda lattice, one can compute the spectrum by using separation of variables \cite{sklyanin}. This amounts to 
quantizing the spectral curve for the two different choices in (\ref{ovals}), obtaining in this way two quantum Baxter equations. 
This method is much faster since it involves two one-dimensional problems, instead of a two-dimensional problem. 

In the case at hand, the procedure goes as follows. We can quantize the curve (\ref{spec-curve}) by simply promoting $p$ to a differential operator in the 
standard way, 
\be
p \rightarrow -\ri \hbar {\rd \over \rd x}, 
\ee
and when acting on a wavefunction $\psi(x)$, one gets
\be
\label{qb-1}
\psi(x+\ri \hbar)+\psi(x-\ri \hbar)+2p_5(\re^{x/2};H_1, H_2) \psi(x)=0.
\ee
If one considers the other sign in (\ref{ovals}), one obtains a second quantum Baxter equation,  
\be
\label{qb-2}
\psi(x+\ri \hbar)+\psi(x-\ri \hbar)-2p_5(\re^{x/2};H_1, H_2) \psi(x)=0.
\ee
The first quantum Baxter equation is precisely the one studied in \cite{cgm}, and its quantization leads to a discrete family of curves in the 
$(H_1, H_2)$ plane which can be actually determined analytically, by using the conjecture put forward in \cite{cgm}. 
The equation (\ref{qb-2}) does not lead to a well-defined real spectrum for arbitrary values of $H_1$ (or $H_2$). However, if one imposes the 
two quantum Baxter equations simultaneously, one finds a 
{\it discrete} set of eigenvalues $(H_1, H_2)$, which agrees with the original spectrum of the quantum integrable system. 
By using this procedure, we have calculated the eigenvalues $(H_1, H_2)$ for many values of $\hbar$. 

The analytic calculation of the spectrum involves a precise determination of the l.h.s. of (\ref{EQC}). 
A simplification occurs for the self-dual value of $\hbar=2\pi$, since 
we can use the simpler quantization condition (\ref{SD}). This only involves the large radius prepotential of the CY and the usual 
mirror map, which can be both obtained from the periods.  As usual in mirror symmetry, we introduce the fundamental period $\varpi_0(\rho_1,\rho_2)$, given by (see \cite{cgm})
\be
\label{fplr} \varpi_0(\rho_1,\rho_2)=\sum_{\ell,n \geq 0}\frac{\Gamma (\rho_1+1)^2 
\Gamma (\rho_2+1) \Gamma (\rho_1-2 \rho_2+1) \Gamma (-3 \rho_1+\rho_2+1)z_1^{\ell+\rho_1}z_2^{k+\rho_2} }
{\Gamma (\ell+\rho_1+1)^2 \Gamma (k+\rho_2+1) \Gamma (\ell-2 k+\rho_1-2 \rho_2+1) \Gamma (-3 \ell+k-3 \rho_1+\rho_2+1)}.
\ee
Let us define, 
\be
\label{genlr-pers}
\ba & {\Pi}_{A_i}={\partial \varpi_0 (\rho_1,\rho_2) \over \partial \rho_i}\big|_{\rho_1=\rho_2=0}, \qquad i=1,2, \\
& {\Pi}_{B_1}= \left(2\partial_{\rho_1}^2+2\partial_{\rho_1} \partial_{\rho_2}+3 \partial_{\rho_2}^2\right)\varpi_0 (\rho_1,\rho_2)\big|_{\rho_1=\rho_2=0},\\
& {\Pi}_{B_2}=\left(\partial_{\rho_1}^2+6 \partial_{\rho_1} \partial_{\rho_2}+9 \partial_{\rho_2}^2\right)\varpi_0 (\rho_1,\rho_2)\big|_{\rho_1=\rho_2=0}.
\ea\ee 
Then, the classical mirror map is given by 
\be
T_i =- {\Pi}_{A_i}(z_1, z_2), \qquad i=1, 2, 
\ee
while the derivatives of the prepotential are given by 
\be
{\partial F_0 \over \partial T_1}= {1\over 10} \Pi_{B_1}, \qquad {\partial F_0 \over \partial T_2}= {1\over 10} \Pi_{B_2}. 
\ee
Using these results, we can already test the quantization condition (\ref{SD}) against numerical calculations of the spectrum. 
In Table \ref{sd-c3z5} we compare the analytic calculation 
of the eigenvalues $H_1$, $H_2$, for the ground state with quantum numbers $(n_1, n_2)=(0,0)$, by using (\ref{SD}), to the numerical result. 
In the analytic calculation, we expand the l.h.s. of (\ref{SD}) up 
to a given order in the $z_i$, $i=1,2$, and we determine the eigenvalues by finding the roots of (\ref{SD}), truncated at that order. 
Clearly, as we increase the order, we approach with higher and higher 
precision the numerical result for the spectrum.

 \begin{table}[t] 
\centering
   \begin{tabular}{l l l}
  \\
Order& $H_1 $  & $H_2$\\
\hline                    
 1 &     \underline{26}.64742838830092 & \underline{37.3}7723389290471\\ 
 10 &   \underline{26.8101214}0803702& \underline{37.3287833}7553935\\ 
 15&    \underline{26.81012141815}522& \underline{37.3287833694}7092\\ 
 20 & \underline{26.81012141815680} &   \underline{37.32878336946952}\\
 \hline
Numerical value &
                             $26.81012141815680$
                        & $ 37.32878336946952$
\end{tabular}
\\
\caption{ The eigenvalues $H_1$ and $H_2$ for the ground state $(n_1,n_2)=(0,0)$ of the $\IC^3/\IZ_5$ GK integrable system, and for $\hbar=2 \pi$, 
as obtained from the quantization 
condition (\ref{SD}). The last line displays the eigenvalues obtained by numerical methods.}
 \label{sd-c3z5}
\end{table}

\begin{center}
 \begin{figure}
\begin{center}
 {\includegraphics[scale=0.35]{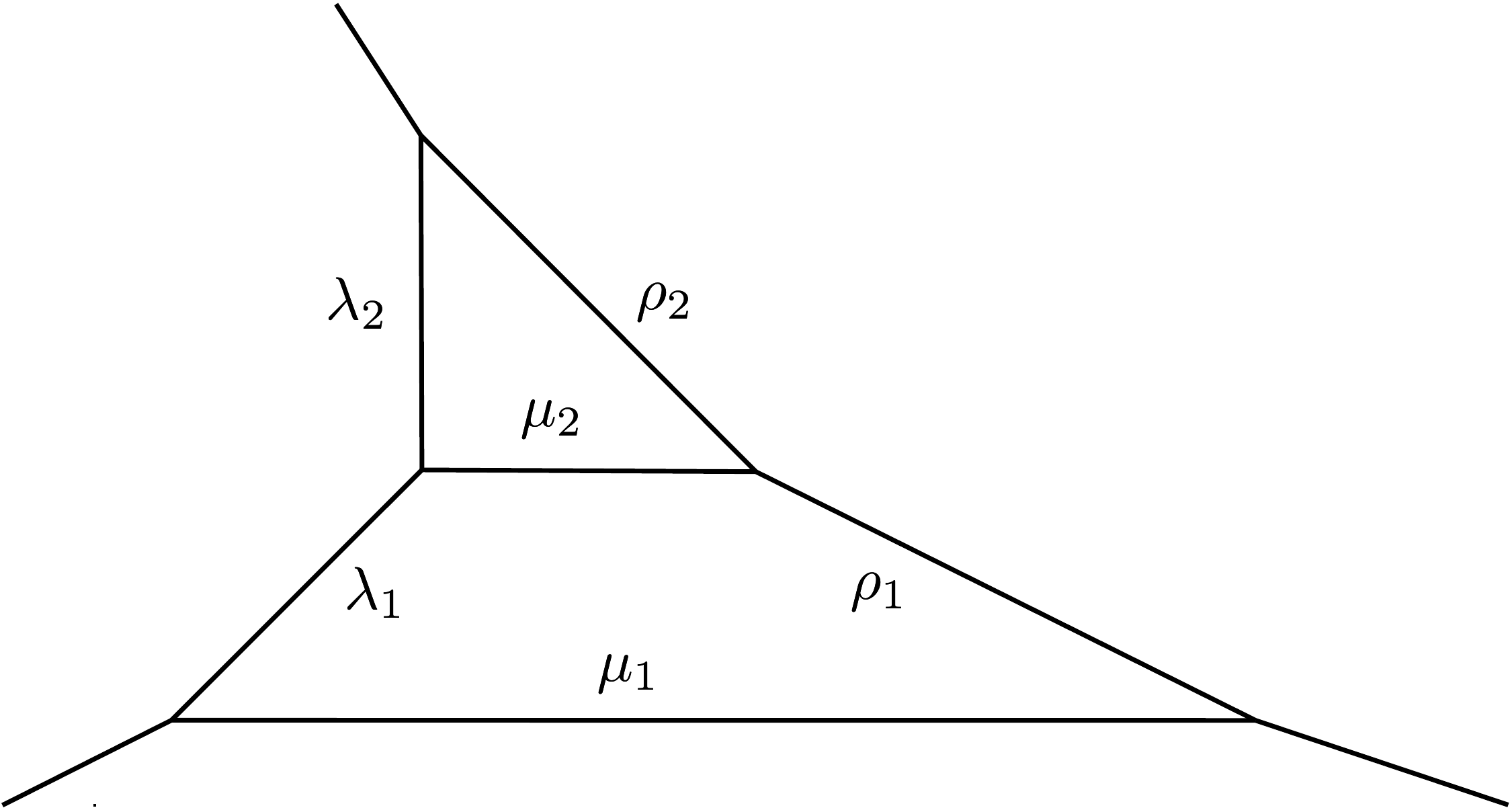}}
\caption{The web diagram for the resolved orbifold $\IC^3/\IZ_5$. The edges are labeled with the 
Young tableaux appearing in the expression (\ref{rvertex}) for the total partition function.}
 \label{web-diag}
\end{center}
\end{figure}  
\end{center}

For general $\hbar$, the analytical calculation requires knowing the NS free energy $F^{\rm NS}$ of the resolved $\IC^3/\IZ_5$ orbifold. 
This can be obtained from the refined topological string partition function, which we compute with
 the refined topological vertex \cite{ikv,ak}. The web diagram for the resolved $\IC^3/\IZ_5$ orbifold
 is shown in \figref{web-diag}. Since the geometry includes a $\IP^2$, we cannot use directly the original refined topological vertex. One possibility is to use the $A_2$ fibration with 
$m=2$ considered in for example \cite{ikp2}, and to perform a flop and a blowdown. Another, equivalent possibility is to 
use the ``new" refined vertex studied in \cite{ik}.  In order to write down the partition function, we need to introduce some notation. 
The refined topological string free energy depends on two parameters $\epsilon_{1,2}$, through the exponentiated variables 
\be
\label{qt}
q=\re^{ \im \epsilon_1},   \qquad t=\re^{-\im \epsilon_2}.
\ee
Given a partition or Young tableau $\mu$, 
we define the quantities
\be
\ba
|\mu|&=\sum_i \mu_i, \\
\Vert \mu \Vert^2&= \sum_i \mu_i^2,\\
\kappa_\mu&= \sum_i \mu_i(\mu_i-2i+1).
\ea
\ee
The refined framing factor is given by 
\be
f_\mu= (-1)^{|\mu|}\left( {t \over q} \right)^{\Vert \mu^t \Vert^2/2} q^{-\kappa_\mu/2}, 
\ee
where $\mu^t$ denotes the transposed partition in which one exchange rows and columns 
of the corresponding Young diagram. The refined topological vertex will be denoted by 
$C_{\lambda \mu \rho}(t,q)$, and we follow the conventions of \cite{ikv,ik}. Quantities where we exchange $t$ and $q$ and hatted, so for example 
\be
\widehat C_{\lambda \mu \rho}(t,q)=C_{\lambda \mu \rho}(q,t). 
\ee
%
%
 \begin{table}[t] 
\centering
   \begin{tabular}{l l l}
  \\
Order& $H_1 $  & $H_2$\\
\hline                                                                           
 1 &     \underline{25.7}4686135537727&   \underline{19.3}1574194214717\\ 
 5 &     \underline{25.78625}560013398 &  \underline{19.33211}445464925\\ 
 10&    \underline{25.7862572552}8139 &  \underline{19.3321156699}4518\\ 
15 &    \underline{25.78625725529248} &  \underline{19.33211566994956}\\
 \hline
Numerical value &
                            $ 25.78625725529248$
                        & $ 19.33211566994956$
\end{tabular}
\\
\caption{ The eigenvalues $H_1$ and $H_2$ of the $\IC^3/\IZ_5$ GK integrable system, for the state with quantum numbers $(n_1,n_2)=(1,0)$ and $\hbar= \pi$, 
as obtained from the quantization 
condition (\ref{EQC}). The order $d$ refers to the total order of the expansion in the variables $Q_2$ and $Q_1^{1/2}$. The last line displays the eigenvalues obtained by numerical methods.}
 \label{c3z5-two}
\end{table}
%
Due to the presence of a $\IP^2$ in the geometry, we need the two-leg specialization of the ``new refined vertex" of 
\cite{ik}, which we will denote by $P_{\lambda \mu}$. This is defined as the coefficient of the highest power of $Q$ in the polynomial 
\be
Z_{\lambda \mu}(Q)= {\sum_{\sigma} (-Q)^{|\sigma|} C_{\lambda^t 0 \sigma} \widehat C_{0 \mu \sigma^t} \over 
\sum_{\sigma} (-Q)^{|\sigma|} C_{0 0 \sigma} \widehat C_{0 0 \sigma^t}}. 
\ee
It is possible to give an expression for $P_{\lambda \mu}$ in terms of Macdonald polynomials, but we will not need it here, see \cite{ik} for 
explanations and references. The final result for the refined partition function of the resolved $\IC^3/\IZ_5$ orbifold is  
\be
\label{rvertex}
\ba
Z\left({\bf Q}, \epsilon_1, \epsilon_2 \right)&=
 \sum_{\mu_1, \mu_2} \sum_{\lambda_1, \lambda_2}\sum_{\rho_1, \rho_2} f_{\mu_1}^{-4} f_{\mu_2}^{-2} f_{\lambda_1 } f_{\lambda_2} \widehat f_{\rho_1}  \widehat f_{\rho_2} 
(-Q_1)^{|\mu_1|+ |\mu_2|+ |\lambda_2|+ |\rho_2|}  (-Q_2)^{3 |\mu_1|+  |\lambda_1|+ |\rho_1|}\\
& \qquad \times \left( {\sqrt{{t\over q}}}\right)^{|\lambda_1|+ |\lambda_2|- |\rho_1| - |\rho_2| } C_{0 \lambda_1 \mu_1} C_{\lambda_1^t \lambda_2 \mu_2} \widehat C_{\rho_1^t 0 \mu_1^t }\widehat C_{\rho_2^t \rho_1 \mu_1^t } P_{\lambda_2 \rho_2}. 
\ea
\ee
The NS free energy is obtained as the limit
\be
F^{\rm NS, \, BPS}({\bf Q}, \hbar) =- \lim_{\epsilon_2 \rightarrow 0} 
\epsilon_2 \log Z ({\bf Q}, \epsilon_1, \epsilon_2), 
\ee
after identifying $\epsilon_1= \hbar$. We find, at the very first orders, 
\be
\ri F^{\rm NS, \, BPS}({\bf Q}, \hbar)=
-{q+ 1+ q^{-1} \over q^{1/2}- q^{-1/2}} Q_1 + {q+1\over q-1} Q_2-{(q^{1/2} + q^{-1/2})^2 \over q^{1/2}- q^{-1/2}} Q_1 Q_2 +  {1\over 4} {q^2+1\over q^2-1} Q^2_2 + \cdots 
\ee
The quantum mirror map of the $\IC^3/\IZ_5$ orbifold can be computed by using the techniques of \cite{acdkv}, as spelled out in detail in \cite{cgm}. Using these results, one can 
calculate explicitly the eigenvalues for different values of $\hbar$. In this case, calculation of the NS free energy at higher orders is computationally expensive. We have pushed the 
calculation to total order $15$ in $Q_2$, $ Q_1^{1/2}$ (the fact that the variables enter in this asymmetric way is due to the 
blowdown procedure we are using in the calculation). We have found excellent agreement for many different values of $\hbar$, providing strong indications that the 
quantization condition (\ref{EQC}) captures the spectrum of the quantum integrable system correctly. We show an example in table \ref{c3z5-two}, for $\hbar=\pi$.

\begin{center}
 \begin{figure}
\begin{center}
 {\includegraphics[scale=0.5]{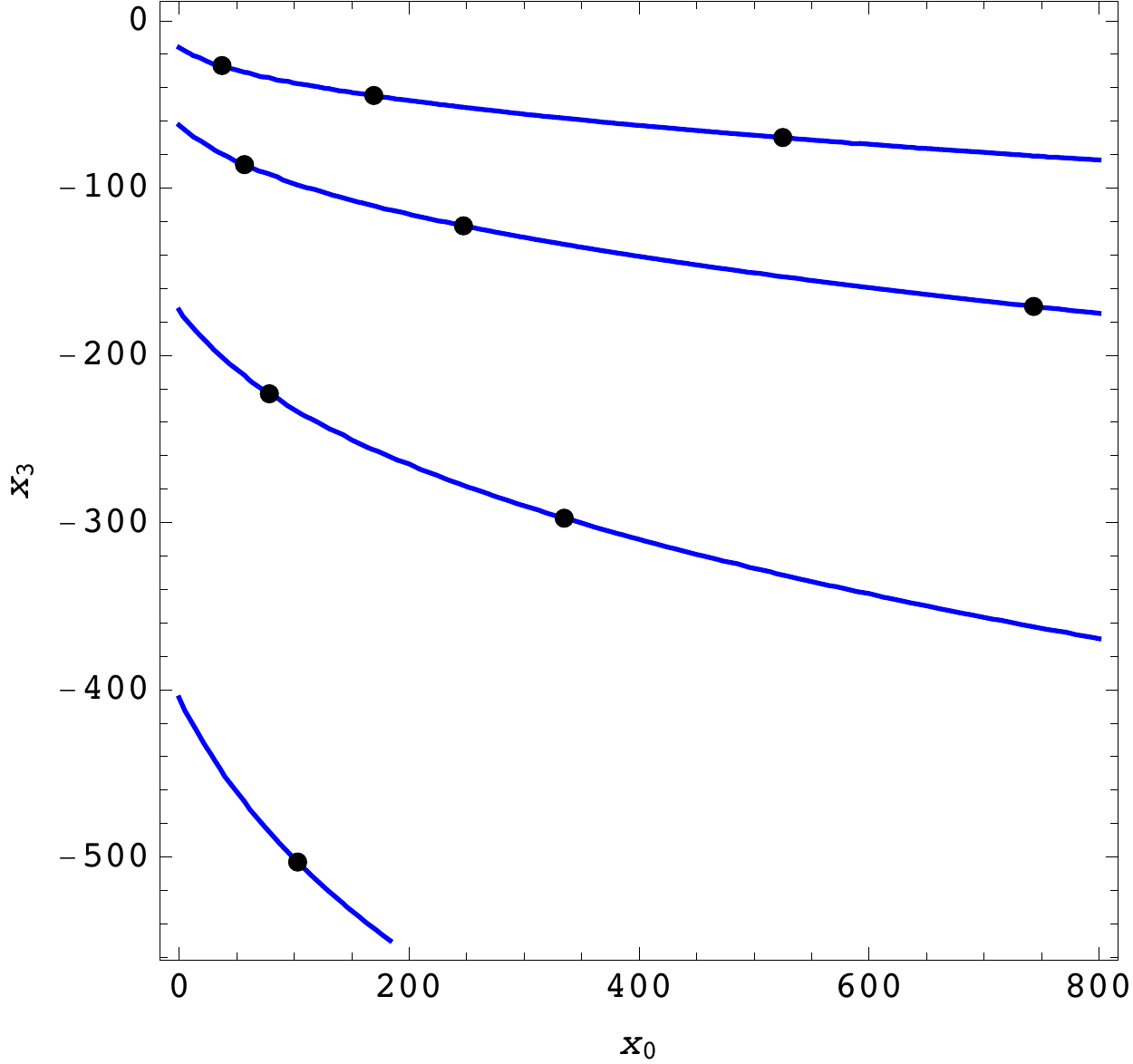}}
\caption{The moduli space of the resolved $\IC^3/\IZ_5$ orbifold has a real slice which is the $(x_0, x_3)$ plane shown in this figure. The curves in this plane are 
defined by the vanishing of the spectral determinant $\Xi(x_0, x_3)$ studied in \cite{cgm}, for $\hbar=2 \pi$. 
The black dots represent the eigenvalues of the 
Hamiltonians $(H_2, -H_1)$ for the very first eigenstates of the corresponding GK quantum integrable system.}
 \label{curve-points}
\end{center}
\end{figure}  
\end{center}

As we explained above, the spectrum of the GK quantum integrable system associated to the resolved $\IC^3/\IZ_5$ orbifold 
can be obtained from the simultaneous consideration of two quantum Baxter equations. The solution to the first one has been 
studied in \cite{cgm}. It is encoded in the zero locus of a generalized spectral determinant $\Xi(x_0, x_3)$. Taking into account the dictionary (\ref{xHs}), we find that the solution to the Baxter equation (\ref{qb-1}) 
is encoded in the quantization condition 
\be
\label{vsd}
\Xi(H_2, -H_1)=0. 
\ee
In \figref{curve-points} we show the eigenvalues of $(H_2, -H_1)$ for the very first bound states of the integrable system, for $\hbar=2 \pi$. As expected, they all lie on the curves 
determined in \cite{cgm}. This can be also verified analytically. Indeed, as shown in \cite{cgm}, in the self-dual case, 
the spectral determinant can be expressed in terms of a genus two theta function. The arguments of this theta function 
are closely related to the functions appearing in the l.h.s. of (\ref{SD}), and it is easy to check that the solutions of (\ref{SD}) also satisfy (\ref{vsd}). 

A natural question is whether the second Baxter equation (\ref{qb-2}) can be solved by a similar quantization condition involving the techniques of \cite{cgm}. 
It can be seen that the change of signs involved in (\ref{qb-2}) can be implemented by a rotation of $x_0$, $x_3$ by a non-trivial phase, so that the quantization condition corresponding to (\ref{qb-2}) 
is given by 
\be
\label{vsd-2}
\Xi\left( \re^{ 6\pi \ri/5} H_2,  \re^{ 3\pi \ri/5} H_1\right)=0. 
\ee
The solutions to this equation are very different from the ones to (\ref{vsd}), since given a real value of $H_1$ (or $H_2$), the corresponding value of $H_2$ (or $H_1$) is complex. We believe that these 
values correspond to resonant states appearing in the solution to the quantum Baxter equation (\ref{qb-2}). By imposing the two quantization conditions (\ref{vsd}) and (\ref{vsd-2}) simultaneously, 
we obtain precisely the spectrum of the GK integrable system associated to the resolved $\IC^3/\IZ_5$ orbifold.

\subsection{The resolved $\IC^3/\IZ_6$ orbifold}

\label{section_Z6_example}

Let us now consider our second example, the resolved $\IC^3/\IZ_6$ orbifold. This is an $A_2$ fibration which was analyzed in the context of geometric engineering in \cite{kkv}, and 
has been recently studied in \cite{klemm-g2}. As it follows from \figref{toric_ZN}, the corresponding Newton polygon has two inner points, leading to a mirror curve of genus two and 
to two mutually commuting Hamiltonians $H_1$, $H_2$. There are four boundary points, therefore one non-trivial Casimir, which is related to $R$, as we saw in section~\ref{section_cluster_ZN}.
On the topological string side, $R$ corresponds to a mass parameter or ``radius." 
The toric geometry of this resolved orbifold can be encoded, as usual, in the vectors of charges
\be
\ba
e_1&= (0,0,0, 1, -2, 1),\\
e_2&=(0,0,1,-2,1,0),\\
e_3&=(1,1,0,0,-2). 
\ea
\ee
Let us write the mirror curve as (see also \cite{bt})
\be
\label{mirror_Z6}
a_1 \re^p+a_2 \re^{-p}+b_3 \re^{3x}+b_2 \re^{2x}+b_1 \re^{x}+b_0=0.
\ee
The coefficients of this curve are related to the Hamiltonians and mass parameter as 
\be
a_1=a_2=\pm R^3, \qquad b_3=1,\qquad b_2=-H_1,\qquad b_1=H_2, \qquad b_0=-H_3
\ee
The Batyrev coordinates for the moduli space are then given by
\be
z_1=\frac{b_0 b_2}{b_1^2}=\frac{H_1H_3}{H_2^2},\qquad
z_2=\frac{b_1 b_3}{b_2^2}=\frac{H_2}{H_1^2}, \qquad
z_3=\frac{a_1 a_2}{b_0^2}=\frac{R^6}{H_3^2}.
\ee
In the center of mass frame ($p_1+p_2+p_3=0$), we can make a reparameterization of \eqref{eq:t-para-even} by
\be
t_1=R^2 \re^{x_1},\quad t_2=\re^{y_1}, \quad t_3=R^2 \re^{-x_1+x_2},\quad
t_4=\re^{y_2},\quad t_5=R^2 \re^{-x_2}, \quad t_6=\re^{-y_1-y_2}.
\label{eq:para-Z6}
\ee
Then the two non-trivial quantum Hamiltonians are now given by
\be
\ba
\mH_1&=R^2(\re^{\mx_1}+\re^{-\mx_2}+\re^{-\mx_1+\mx_2})+\re^{\my_1}+\re^{\my_2}+\re^{-\my_1-\my_2}, \\
\mH_2&=R^4(\re^{-\mx_1}+\re^{\mx_2}+\re^{\mx_1-\mx_2})+R^2(\re^{\mx_1+\my_2}+\re^{-\mx_2+\my_1}
\re^{-\mx_1+\mx_2-\my_1-\my_2}) \\
&\quad +\re^{-\my_1}+\re^{-\my_2}+\re^{\my_1+\my_2} .
\ea
\label{eq:H-Z6}
\ee
The third Hamiltonian is not dynamical, and should be regarded as a Casimir,  as explained in section~\ref{section_cluster_ZN}:
\be
H_3=1+R^6.
\ee
Thus the Batyrev coordinates are finally given by
\be
z_1=(1+R^6) \frac{H_1}{H_2^2},\qquad
z_2=\frac{H_2}{H_1^2}, \qquad
z_3=\frac{R^6}{(1+R^6)^2}.
\label{eq:moduli-Z6}
\ee
As in the previous example, it is straightforward to connect (\ref{mirror_Z6}) to the general Newton polygon in \figref{toric_ZN}. 
With this map, we can express $z_3$ in terms of the Casimirs (\ref{C1_C2}) and (\ref{C3}) as follows
\beq
z_3={C_1 C_2\over C_3^2}=\frac{t_1 t_2 t_3 t_4 t_5 t_6}{(t_1 t_3 t_5+t_2 t_4 t_6)^2} \, .
\eeq
The other geometric data of the CY are the following. The matrix $\CC_{kj}$ appearing 
in (\ref{EQC}) is given by \cite{klemm-g2}
\be
\label{C-ma-2t}
\CC=\begin{pmatrix} 2 & -1\\ -1 &2 \\ 0& 0 \end{pmatrix}. 
\ee
The perturbative part of the NS free energy can be read from the perturbative prepotential and from the coefficients $b_i^{\rm NS}$, $i=1,2$, which have been computed in \cite{klemm-g2}. One finds, 
\be
\ba
\hbar F^{\rm NS,\, pert }({\bf T}, \hbar)&= {2\over 9} T_1^3 +{1\over 3} T_1^2 T_2+{1\over 3} T_1^2 T_3 +{2\over 3} T_1 T_2^2 +{1\over 3} T_1 T_2 T_3+ {1\over 3} T_2^2 T_3+ 
{4\over 9}T_2^3\\
&- {4 \pi^2 + \hbar^2 \over 6} \left( T_1+T_2  \right). 
\ea
\ee
The B-field for this geometry vanishes. 

The study of our exact quantization conditions for this model is very similar to the previous one. A numerical calculation of the spectrum can be performed either by 
diagonalizing the original Hamiltonians \eqref{eq:H-Z6}, or by studying the two quantum Baxter equations obtained from the two realizations of the spectral curve, 
\be
R^3\left(\psi(x+ \ri \hbar) +\psi(x- \ri \hbar)\right)\pm 2p_6(\re^{x};H_1,H_2)=0,
\ee
where
\be
2 p_6(z; H_1, H_2)=z^3-H_1 z^2+H_2 z-1-R^6.
\ee
In the self-dual case $\hbar= 2\pi$, our quantization condition requires only the special geometry of the CY. The fundamental period at large radius is
\be
\omega_0( \boldsymbol{\rho})=\sum_{l,m,n\ge 0} c(l,m,n; \boldsymbol{\rho})z_1^{k+ \rho_1} z_2^{l+ \rho_2} z_3^{m+ \rho_3},
\ee
where
\be
\ba
c(k,l,m; \boldsymbol{\rho})&={1\over \Gamma (m+\rho_3+1)^2 \Gamma (l+\rho_2+1) 
\Gamma(k-2m+\rho_1-2\rho_3+1)} \\
    & \times{1\over  \Gamma (k-2 l+\rho_1-2 \rho_2+1) \Gamma (-2 k+ l-2 \rho_1+\rho_2+1)}.
   \ea
 \ee
We define, 
\be
\label{genlr-pers-2}
\ba & {\Pi}_{A_i}={\partial \varpi_0 (\boldsymbol{\rho}) \over \partial \rho_i}\big|_{\boldsymbol{\rho}=0}, \qquad i=1,2,3, \\
& {\Pi}_{B_1}= \left(2\partial_{\rho_1}^2+2\partial_{\rho_1} \partial_{\rho_3}+2\partial_{\rho_1} \partial_{\rho_2}+\partial_{\rho_2} \partial_{\rho_3
}+2 \partial_{\rho_2}^2\right)\varpi_0 (\boldsymbol{\rho})\big|_{\boldsymbol{\rho}=0},\\
& {\Pi}_{B_2}=\left( \partial_{\rho_1}^2+\partial_{\rho_1} \partial_{\rho_3}+4\partial_{\rho_1} \partial_{\rho_2}+2\partial_{\rho_2} \partial_{\rho_3
}+4\partial_{\rho_2}^2\right)\varpi_0 (\boldsymbol{\rho})\big|_{\boldsymbol{\rho}=0}.
\ea\ee 
Then, the mirror map is given by 
\be
T_i =- {\Pi}_{A_i}(z_1, z_2, z_3), \qquad i=1, 2, 3.  
\ee
Since $T_3$ corresponds to a mass parameter, its mirror map is algebraic, 
\be
T_3=\log(z_3) - 2 \log\left( {1-{\sqrt{1- 4z_3}} \over 2} \right). 
\ee
This mirror map, together with \eqref{eq:moduli-Z6}, shows that the radius $R$ is given by 
\be
Q_3=\re^{-T_3}=R^6.
\ee
The derivatives of the prepotential are given by 
\be
{\partial F_0 \over \partial T_1}= {1\over 3} \Pi_{B_1}, \qquad {\partial F_0 \over \partial T_2}= {1\over 3} \Pi_{B_2}. 
\ee
Due to the presence of the additional parameter $z_3$, the convergence of the large radius expansion is slower than in the previous example. However, 
we can use the results on the special geometry of this CY threefold to produce highly non-trivial tests of the conjecture in the self-dual case. An example is shown in Table \ref{c3z6-sd}, for $R=1/2$.

 \begin{table}[t] 
\centering
   \begin{tabular}{l l l}
  \\
Order& $H_1 $  & $H_2$\\
\hline                    
 10 &  \underline{17.2598}2433969  &  \underline{29.41243}745481\\ 
 15&   \underline{17.2598321}5076  &  \underline{29.4124369}8921\\ 
 20 &  \underline{17.259832196}19  &  \underline{29.4124369911}5\\
  25 & \underline{17.25983219649}  &  \underline{29.41243699119}\\
 \hline
Numerical value &
                             $17.25983219649$
                        &   $29.41243699119$
\end{tabular}
\\
\caption{ The eigenvalues of $H_1$ and $H_2$ for the ground state $(n_1,n_2)=(0,0)$ of the $\IC^3/\IZ_6$ GK integrable system, as obtained from the quantization 
condition (\ref{SD}). We set $\hbar=2 \pi$ and $R=1/2$. The last line displays the eigenvalues obtained by numerical methods.}
 \label{c3z6-sd}
\end{table}

When $\hbar\not=2 \pi$, we have to compute the NS free energy of this CY. Since the resolved $\IC^3/\IZ_6$ orbifold is the $A_2$ fibration with 
Chern--Simons number $m=3$ (see for example \cite{ikp2}), we can use 
the formula for the refined partition function with arbitrary $m$ written down in \cite{hm} (following \cite{ikv,taki}.) 
We have performed the computation of the NS free energy up to order $12$.
The very first few terms are given by
\be
\ri F^{\rm NS, \, BPS}({\bf Q}, \hbar)=
\frac{q+1}{q-1}(Q_1+Q_2)+\frac{2q}{q^2-1}Q_1(Q_2+Q_3)
+\frac{q^2+1}{4(q^2-1)}(Q_1^2+Q_2^2+4Q_1Q_2+4Q_1Q_3)+ \cdots 
\ee
The quantum mirror map of this geometry can be also 
calculated with the techniques of \cite{acdkv} (details of this calculation will appear elsewhere). 
Using these data, we can evaluate the eigenvalues for generic values of $\hbar$.
In Tables~\ref{c3z6-pi}, \ref{c3z6-3}, we show the eigenvalues of an excited state $(n_1,n_2)=(0,1)$ 
for $\hbar=\pi$ and $\hbar=3$ with $R=1/2$, respectively.
As in the previous example, the results obtained from our quantization condition converge to
the right eigenvalues as higher order corrections are included.

 \begin{table}[t] 
\centering
   \begin{tabular}{l l l}
  \\
Order& $H_1 $  & $H_2$\\
\hline                    
 1 &   \underline{14.6}704208685885 &  14.9636742542436\\ 
 4 &   \underline{14.68278}09718360 &  14.9796593227273\\ 
 8 &   \underline{14.6827811605}017 &  14.9796598251795\\
 12 & \underline{14.6827811605316}  &  \underline{14.9796598252955}\\
 \hline
Numerical value &
                             $14.6827811605316$
                        &   $14.9796598252955$
\end{tabular}
\\
\caption{ The eigenvalues of $H_1$ and $H_2$ for the excited state $(n_1,n_2)=(0,1)$ of the $\IC^3/\IZ_6$ GK integrable system, as obtained from the quantization 
condition (\ref{EQC}). We set $\hbar=\pi$ and $R=1/2$. The last line displays the eigenvalues obtained by numerical methods.}
 \label{c3z6-pi}
\end{table}

 \begin{table}[t] 
\centering
   \begin{tabular}{l l l}
  \\
Order& $H_1 $  & $H_2$\\
\hline                    
 1 &   \underline{13.8}6846367370  &  \underline{14.1}2507965160\\ 
 4 &   \underline{13.879567}58865  &  \underline{14.13905}792133\\ 
 8 &   \underline{13.8795674659}1  &  \underline{14.139059290}02\\
 12 & \underline{13.87956746586}  &  \underline{14.13905929013}\\
 \hline
Numerical value &
                             $13.87956746586$
                        &   $14.13905929013$
\end{tabular}
\\
\caption{ The eigenvalues of $H_1$ and $H_2$ for the excited state $(n_1,n_2)=(0,1)$ of the $\IC^3/\IZ_6$ GK integrable system, as obtained from the quantization 
condition (\ref{EQC}). We set $\hbar=3$ and $R=1/2$. The last line displays the eigenvalues obtained by numerical methods.}
 \label{c3z6-3}
\end{table}

\sectiono{Conclusions and open problems}

\label{section_conclusions}

In this paper we have proposed an exact quantization condition for the quantum integrable system of \cite{gk}. This involves a perturbative part, given by the NS limit of the 
refined topological string free energy, and a non-perturbative part. The non-perturbative contribution is presumably due to complex instantons \cite{km}, but as noted in \cite{wzh,hatsuda} 
in the genus one case, it can be related to the perturbative part by a simple S-duality symmetry. Our proposal generalizes the conjecture for the relativistic Toda lattice put forward in \cite{hm}, 
and incorporates in a crucial way the recent progress on quantization of mirror curves \cite{km,hw,ghm, kas-mar,mz,kmz,wzh,gkmr}

Of course, the main open issue is to prove (or disprove) our conjecture. In this paper we have provided what we find is compelling empirical evidence for its 
validity. Note that, in the case of curves of genus one, our proposal gives precisely the conjectural quantization conditions of \cite{ghm}, in the form obatained in \cite{wzh}. 
Therefore, the extensive evidence in favor of \cite{ghm} can be regarded as additional support for the conjecture put forward in this paper. Another obvious question is to find 
the eigenfunctions corresponding to the eigenvalues determined by our conjectural quantization condition. These will probably involve topological open string amplitudes on $X_N$.  

As we have pointed out in this paper, it would be very interesting to further clarify the relationship between the quantization of the GK system, and the quantization of higher genus curves 
proposed in \cite{cgm}. One appealing feature of the framework put forward in \cite{cgm} is that the conventional topological string free energy of the CY is fully reconstructed 
by a 't Hooft-like limit of the spectral traces for the appropriate operators. An intriguing question is whether the information on the exact quantization conditions for the 
GK system can be reformulated in terms of a Fredholm determinant or in terms 
of appropriate spectral traces, as in \cite{cgm}.

\section*{Acknowledgements}
 We would like to thank Andrea Brini, Santiago Codesido, Vladimir Fock, Alba Grassi, Jie Gu, Albrecht Klemm and Szabolcs Zakany for discussions. The work of S. F. is supported by the U.S. National Science 
 Foundation grant PHY-1518967 and by a PSC-CUNY award. The work of Y.H. and M.M. is supported in part by the Fonds National Suisse, 
subsidies 200021-156995 and 200020-159581, and by the NCCR 51NF40-141869 ``The Mathematics of Physics" (SwissMAP). 

\appendix

\end{document}